\documentclass[aps,eqsecnum,twocolumn]{revtex4}
\usepackage{epsfig}
\newif\ifmgf
\mgftrue
\newcommand{\vv}[1]{{\mathbf{#1}}}
\newcommand{\bs}[1]{\mbox{\boldmath $#1$}}
\newcommand{\tr}{{\mathrm{tr}}}
\newcommand{\T}{{\mathrm{T}}}

\begin{document}

\title{Robust statistics for deterministic and stochastic\\
gravitational waves in non-Gaussian noise I: Frequentist analyses.} 

\author{Bruce Allen and Jolien D. E. Creighton}
\affiliation{Department of Physics, University of Wisconsin - Milwaukee,
P.O. Box 413, Milwaukee WI 53201}

\author{\'Eanna \'E.\ Flanagan}
\affiliation{Newman Laboratory of Nuclear Studies, Cornell University,
Ithaca, NY 14853-5001}

\author{Joseph D. Romano}
\affiliation{Department of Physical Sciences, University of
Texas at Brownsville, Brownsville TX 78520}

\begin{abstract}\quad
Gravitational wave detectors will need optimal signal-processing
algorithms to extract weak signals from the detector noise.  Most
algorithms designed to date are based on the unrealistic assumption
that the detector noise may be modeled as a stationary Gaussian
process.  However most experiments exhibit a non-Gaussian ``tail'' in
the probability distribution.  This ``excess'' of large signals can be
a troublesome source of false alarms.  This article derives an optimal
(in the Neyman-Pearson sense, for weak signals) signal processing
strategy when the detector noise is non-Gaussian and exhibits tail
terms.  This strategy is robust, meaning that it is close to optimal
for Gaussian noise but far less sensitive than conventional methods to
the excess large events that form the tail of the distribution.  The
method is analyzed for two different signal analysis problems: (i) a
known waveform (e.g., a binary inspiral chirp) and (ii) a stochastic
background, which requires a multi-detector signal processing
algorithm.  The methods should be easy to implement: they amount to
truncation or clipping of sample values which lie in the outlier part
of the probability distribution.
\end{abstract}

\maketitle


\section{INTRODUCTION}
\label{s:intro}
The construction of several new detectors of gravitational radiation is
currently approaching completion.  These instruments are of a different design
and have significantly better sensitivity and broader bandwidth than previous
detectors.  They include the LIGO detector being built in the United States by
a joint Caltech/MIT collaboration \cite{science92,physicstoday}, the VIRGO
detector being built near Pisa by an Italian/French collaboration
\cite{virgo}, the GEO-600 detector being built in Hannover by an Anglo/German
collaboration \cite{geo600}, and the TAMA-300 detector near Tokyo
\cite{tama300}.  There are also several resonant bar detectors currently in
operation \cite{barsinoperation}, and several more refined bar and
interferometric detectors presently in the planning and proposal stages
\cite{plannedinstruments}.  These instruments search for very weak signals.
For the most likely sources, the signals will be buried in the noise of the
detectors, and need to be extracted with sophisticated optimal
signal-processing strategies \cite{300years}.

The standard assumption made in the literature is that the detector noise has
multivariate Gaussian statistics.  This assumption is certainly incorrect:
every sensitive gravitational wave detector operated to date has been
characterized by noise that is both non-stationary and non-Gaussian.  Some
experimentation has shown that this is a serious matter \cite{exptwith40m}:
existing detection strategies for both deterministic and stochastic signals do
not perform nearly as well when non-Gaussian noise is present.  Roughly
speaking, if the non-Gaussian fluctuations are large, they bias the statistics
and make it more difficult to achieve a given level of statistical confidence.

In this paper, we develop a new set of statistical signal-processing
techniques to search for deterministic and stochastic gravitational waves.
These techniques are \emph{robust}, meaning that they will work well even if
the detector noise is not Gaussian but falls into a broader statistical class
that we expect includes realistic detectors.  In large part, these new methods
are similar to the older ones: one constructs matched filters to search 
for known
waveforms or cross-correlates the instrument outputs at the different detector
sites to search for a stochastic background.  The essential difference is that
by using locally optimal methods \cite{kassam} these statistical measures are
modified.  The effect is to {\it truncate} the statistics: detector samples
that fall outside the central Gaussian-like part of the sample distribution
(i.e., the outliers) are excluded from (or saturated when constructing) the
measurement statistic.  For both cases, a robust statistic is found which
performs better than the optimal linear filter in the case where the detector
noise is non-Gaussian, and almost as well in the Gaussian-noise case.

The paper is organized as follows.  In Sec.~\ref{s:deterministic} we derive
the locally most powerful signal-processing tests for deterministic signals.
We begin in Sec.~\ref{ss:sdwnkw} with a derivation of the Neyman-Pearson
criteria for optimality, in the case where a known waveform is hidden in white
noise.  We define the power function of a test and derive a criteria for the
locally optimal test in the weak-signal regime.  The locally optimal test is
analyzed for a number of different types of non-Gaussian noise, and we show
that the locally optimal decision statistic is a matched filter where the
non-Gaussian sample values are truncated or excluded.  In Sec.~\ref{ss:sdcnkw}
the results are generalized to the case where a known waveform is hidden in
colored noise, and we introduce models for non-Gaussian colored noise. In
Sec.~\ref{s:stochastic}, we turn to the detection of a stochastic background.
Section~\ref{ss:2dwnss} considers the case of a stochastic signal (i.e., where
the waveform is not known) and derives the locally optimal statistic which can
be used to correlate two identical detectors, where we assume that each
detector has independent white noise and is co-aligned and coincident.  In
Sec.~\ref{ss:2dcnss}, these results are generalized to the case where 
the noise is colored, and the detectors are in different
locations, and not aligned in parallel.  
In Sec.~\ref{s:impsim}, we discuss an implementation of these
statistics, and we illustrate how one can compare the performance of
different statistics using Monte Carlo simulations.
Section~\ref{s:conclude} contains a short conclusion and summary.

\section{DETERMINISTIC SIGNALS}
\label{s:deterministic}

\subsection{Single detector, white noise}
\label{ss:sdwnkw}
In order to describe the idea in a simple way we first discuss the case where
we are searching for a known signal in the data stream of a single detector,
where the time-domain detector noise samples are independent in the
time-domain.

Denote the data stream of the first detector by $\vv{x}_1=x_{1,j}$ for
$j=0,\ldots,N-1$.  In this section, since we are going to only consider this
single detector, we will drop the subscript ``1.'' Imagine that we are looking
for a signal of known waveform but unknown amplitude $\epsilon$, which we will
denote by $\epsilon s_j$.  Our primary interest is in the case where the
amplitude $\epsilon$ is either small, or zero.  For convenience, imagine for
the moment that this parameter can have only two possible values, either
$\epsilon=0$ or $\epsilon = \bar \epsilon \ne 0$.

The detection problem that we need to solve is to partition the space of
possible observations $R^N$ into two disjoint subsets.  When the observation
$\vv{x}$ falls into one of these, we conclude that $\epsilon=0$ and that the
null hypothesis is true.  When the observation falls into the other set we
conclude that the signal has been observed with $\epsilon\ne0$.  To describe
the partition of $R^N$ into two regions, define a function
$\delta(\vv{x}\in R^N)$ which is zero in the null hypothesis region and unity
elsewhere.  This function is called a \emph{test}.  Our goal is to find the
``best'' choice of a test $\delta$.

To help characterize tests $\delta$, it is helpful to define the \emph{power}
function of a test:
\begin{equation}
  F(\delta|\epsilon) = \int_{R^N} \delta(\vv{x}) p(\vv{x}|\epsilon) d^N x.
\end{equation}
Here $p(\vv{x}|\epsilon)$ is the probability distribution of the measurement
$\vv{x}$ given signal amplitude $\epsilon$.  For example, for additive white,
stationary Gaussian noise of unit variance and vanishing mean,
\begin{equation}
\label{e:gaussnoise}
  p(\vv{x}|\epsilon)=\prod_{i=0}^{N-1}(2\pi)^{-1/2}e^{-(x_i-\epsilon s_i)^2/2}.
\end{equation}
The quality of the test can be expressed in terms of the power function.

We characterize the quality of the test by the false alarm and the false
dismissal probabilities.  The false alarm probability is the probability with
which we conclude that $\epsilon\ne0$ when in fact $\epsilon$ vanishes.  
This is given
by $F(\delta|0)$.  The false dismissal probability is the probability with
which we conclude that $\epsilon=0$ when in fact it is
$\epsilon=\bar\epsilon\ne0$.  This is given by $1-F(\delta|\bar \epsilon)$.

One standard definition of the ``best'' test $\delta$ is that it minimizes the
false dismissal probability for a given false alarm probability.  This is
called the Neyman-Pearson test.  One can find this test using calculus of
variations, with a Lagrange multiplier $\Lambda$ to enforce the constraint
that the false alarm probability is fixed.  The best test is obtained by
partitioning $R^N$ as follows.  Choose a constant $\Lambda_0>0$. Then, set
$\delta=1$ in regions where the likelihood ratio
\begin{equation}
\label{e:like}
  \Lambda = \frac{p(\vv{x}|\bar\epsilon)}{p(\vv{x}|0) }
\end{equation}
is greater than $\Lambda_0$.  Set $\delta=0$ elsewhere. (We assume that the
boundary between these two regions is a set of probability measure zero.)  The
value of the constant $\Lambda_0$ determines the false alarm probability.
Thus, the likelihood ratio is a ``decision statistic'': a number that can be
calculated from the observed data.  If the statistic is less than some value,
we conclude that the null hypothesis holds.  If the statistic is greater than
this value, we conclude the opposite. The decision statistic provides a
partition of the space of observations into two disjoint regions.

In the case where the noise is Gaussian (\ref{e:gaussnoise}) this criteria is
easily understood.  The optimal Neyman-Pearson test divides the space of
observation along an $(N-1)$-dimensional plane.  
On one side of this plane $\delta=1$ and on
the other side $\delta$ vanishes.  The plane is defined by setting the
likelihood ratio (\ref{e:like}) to a constant.  For the Gaussian probability
distribution (\ref{e:gaussnoise}) the plane is defined by
\begin{eqnarray}
  \text{constant} & = &
  \prod_{i=0}^{N-1} e^{-(x_i - \epsilon s_i)^2/2 + x_i^2/2}
  \Rightarrow \cr
  \text{constant}& = & \epsilon \sum_{i=0}^{N-1}  x_i  s_i.
\label{e:simplegauss}
\end{eqnarray}
This plane is perpendicular to the vector $\vv{s}$.  Different choices of this
plane correspond to different false alarm rates.

\begin{figure}[t]
\epsfig{file=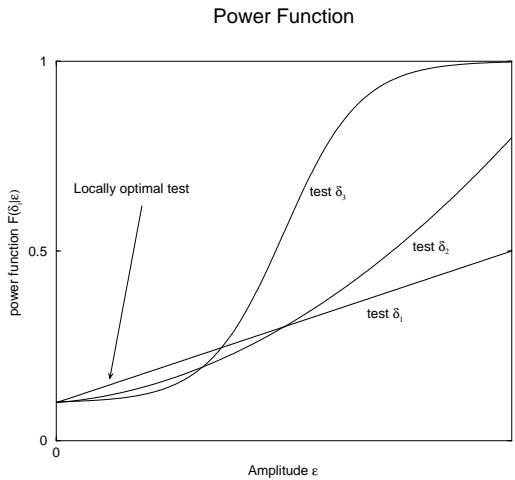,height=7cm,angle=-90}
\caption{
\label{f:powerfn}
The power function $F(\delta|\epsilon)$ is shown for three different tests.
All have the same false alarm probability $F(\delta|0)$.  Test $\delta_3$ has
the best performance for large $\epsilon$.  Test $\delta_2$ is not the best
test for any value of $\epsilon$.  Test $\delta_1$ is the best test for small
$\epsilon$.  The \emph{locally optimal} test $\delta_1$ is the one for which
$dF(\delta|\epsilon)/d\epsilon$ is largest at $\epsilon=0$.  If the first
derivative of the power function with respect to $\epsilon$ vanishes for all
tests, then the locally optimal test is the one with the largest second
derivative (and so on, if additional derivatives vanish).}
\end{figure}

In the case where the noise is not Gaussian, the problem becomes more
challenging.  In the Gaussian test, the decision statistic is independent of
the signal amplitude $\epsilon$. However when the noise in not Gaussian, the
choice of decision statistic depends upon $\epsilon$.  Consider the graph in
Fig.~\ref{f:powerfn} showing the power function $F(\delta|\epsilon)$ as a
function of $\epsilon$ for several different tests. All the tests have the
same false alarm rate, but the optimal test depends upon the value of
$\epsilon$.

For the case of weak signals in non-Gaussian noise, there is a useful test
called the ``locally optimal'' test. For a given noise probability
distribution, the locally optimal test is easy to describe, and leads to a
simple decision statistic which can be calculated from the observed data
\cite{kassam}.  To define this test, it is useful to again consider the set of
all tests with a given false alarm rate, as shown for example in
Fig.~\ref{f:powerfn}.  The locally optimal test is the one that maximizes
$dF(\delta|\epsilon)/d\epsilon$ at $\epsilon=0$ for a fixed false alarm
probability.  As above, one can show that the locally optimal test sets
$\delta=1$ inside the region where
\begin{equation}
\label{e:lotest}
\Lambda_{(1)}=[d\ln p(\vv{x}|\epsilon)/d\epsilon]_{\epsilon=0}>\text{constant}
\end{equation}
for some constant, and $\delta=0$ elsewhere (see Fig.~\ref{f:powerfn}).  The
value of the constant determines (or is determined by) the false alarm
probability.  More generally, if the first derivative vanishes, the locally
optimal test is determined by the first non-zero
\begin{equation}
\label{e:lotest2}
  \Lambda_{(n)}=\left.
  \frac{1}{p(\vv{x}|0)}\frac{d^n p(\vv{x}|\epsilon)}{d\epsilon^n}
  \right|_{\epsilon=0}.
\end{equation}
To understand the implications of this, it is helpful to consider several
examples.

\begin{figure}
\epsfig{file=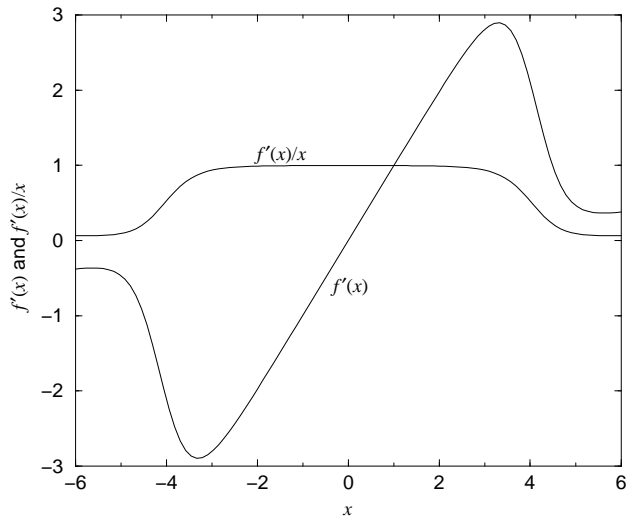,height=7cm}
\caption{
\label{f:weight2}
The functions $f'(x)$ and $f'(x)/x$ are shown for the sum of distinct Gaussian
processes, defined by Eq.~(\protect{\ref{sgs}}) with parameters $\sigma=1$,
$\bar\sigma=4$, and $P=1\%$.  For small $|x|$ one has $f'(x) \approx x$.
Outside the central Gaussian region (which dominates the probability density),
i.e., for large $|x|$, $f'(x)$ falls off. This effectively ``clips'' the
correlation statistic for outlier data samples.}
\end{figure}

The examples here are for the case where the (additive) detector noise is
independent for each sample value (so the noise spectrum is white) but has an
arbitrary probability distribution. For convenience, we write
\begin{equation}
\label{e:nongaussnoise}
  p(\vv{x}|\epsilon) = \prod_{i=0}^{N-1} e^{-f(x_i-\epsilon s_i)}
\end{equation}
where the function $f$ is a quadratic function of its argument for the case
where the probability distribution of the noise is Gaussian.  [Note: any
probability distribution for stationary additive noise where the sample values
are independent can be written in this way. If the noise is not stationary but
is still additive and independent, then each function $f$ appearing in
(\ref{e:nongaussnoise}) may be different $f(x_i-\epsilon s_i) \rightarrow 
f_i(x_i-\epsilon s_i)$.]  The
first derivative of the PDF (\ref{e:nongaussnoise}) with respect to $\epsilon$
is
\begin{equation}
  \frac{d\ln p(\vv{x}|\epsilon)}{d\epsilon}
  = \sum_{i=0}^{N-1}s_i f'(x_i-\epsilon s_i)
\end{equation}
where $f'$ denotes the derivative of $f$ with respect to its argument.
Setting $\epsilon=0$ in this expression one can easily find the locally
optimal test (\ref{e:lotest}).  This is defined by setting $\delta=1$ in the
region
\begin{equation}
\label{e:forexamples}
  \Lambda_{(1)} = \sum_{i=0}^{N-1} s_i f'(x_i) >  \text{constant}
\end{equation}
and setting $\delta=0$ elsewhere.  [Note: if $\epsilon$ can take either sign
$\pm \bar \epsilon$ then an absolute value sign should enclose the LHS of the
inequality in Eq.~(\ref{e:forexamples}).]  As before, the value of the
constant determines the false alarm probability.  Here are several examples:
\begin{itemize}
\item
\textbf{Gaussian Noise}: $f(x) = x^2/2 + \ln (2 \pi)/2$, so $f'(x) = x$.  For
this case the locally optimal test (\ref{e:forexamples}) and the optimal test
(\ref{e:simplegauss}) both give the same statistic: $\sum_{i=0}^{N-1} s_i
x_i$. This is the standard optimal linear filter.
\item
\textbf{Exponential Noise}: $f(x) = a |x| - \ln(a/2)$, so $f'(x) =
a{\mathrm{sgn}}(x)$. Here the locally optimal statistic is given by
(\ref{e:forexamples}) as
\begin{equation}
\label{e:exp}
  \sum_{i=0}^{N-1} s_i {\mathrm{sgn}}(x_i)
\end{equation}
where the ${\mathrm{sgn}}(x)$ function is $+1$ for $x\ge0$ and $-1$ for $x<0$.
\item
\textbf{Sum of distinct Gaussian processes}: this is a white-noise version of
the model given in~\cite{creighton99}:
\begin{eqnarray}
\label{sgs}
  e^{-f(x)} &=& (1-P)(2\pi)^{-1/2}\sigma^{-1} e^{-x^2/2\sigma^2} \nonumber\\
  && + P (2\pi)^{-1/2} \bar\sigma^{-1} e^{-x^2/2\bar\sigma^2},
\end{eqnarray}
where $0<\sigma<\bar\sigma$ and $P\in(0,1)$.  Usually one also has $P\ll1$.
This noise model is discussed in more detail later in this paper. It often
arises when the most common source of noise is Gaussian, but there is also a
``tail'' of ``outlier'' events which dominates the wings of the distribution.
Here the locally optimal statistic is defined by (\ref{e:forexamples}) where
\[
f'(x) = x\sigma^{-2}\left[
\frac{(1-P) + P(\sigma/\bar\sigma)^3 e^{x^2(\sigma^{-2}-\bar\sigma^{-2})/2}}
     {(1-P) + P(\sigma/\bar\sigma) e^{x^2(\sigma^{-2}-\bar\sigma^{-2})/2}}
\right].
\]
This function is shown (for the case $\sigma=1$, $\bar\sigma=4$, $P=1\%$)
in Fig.~\ref{f:weight2}.  Roughly speaking, for $|x|$ small compared to
$\sigma$ one has $f'(x)\approx x/\sigma^2$. For large $|x|$ one has
$f'(x)\approx x/\bar\sigma^2$.
\item
\textbf{Gaussian noise plus uniform background}: Here, we have a (small)
uniform background superposed on Gaussian noise of zero mean and unit
variance.  This is defined for (small) $P>0$ by
\begin{equation}
\nonumber
e^{-f(x)} = \left\{
\begin{array}{ll}
  (1-P)(2\pi)^{-1/2}e^{-x^2/2}+P/2L, & |x|\le L \\
  0, & |x|>L
\end{array} \right.
\end{equation}
Here we assume that $L\gg1$ is the scale size of the uniform background (the
probability distribution is correctly normalized only in the limit
$L\to\infty$).  In this case one finds that $f'(x)\approx x$ for
$|x|\lesssim1$ and $f'(x)=0$ for $1\lesssim|x|\le L$.
\end{itemize}
While the results for the different probability distributions are technically
different, they all carry same message, which is the central result of this
paper: \emph{If the distribution of sample values has a central Gaussian
region, then sample values falling in this region should be correlated exactly
as they would be in the Gaussian case.  If a sample value falls outside this
region, its value should be truncated (or clipped) to the largest allowed
value in the central region, or even dropped from any correlation statistic,
depending upon the shape of the probability distribution}.

Let us repeat this central point one more time.  The results show that when
the noise is not Gaussian, the normal optimal filter used to construct a
decision statistic is replaced by a somewhat different sum.  The values of the
expected signal $s_i$ are multiplied, not by the observed data $x_i$ but by
some non-linear function of that data, then summed.  In the event that the
probability distribution of the noise has a non-Gaussian tail, the effect of
this non-linear function is to ``clip'' or truncate sample values which fall
outside the central bulge of the probability distribution function.

\subsection{Single detector, colored noise}
\label{ss:sdcnkw}
If the detector's noise spectrum is colored rather than white, then the
previous analysis does not apply: the assumption that the different sample
values are uncorrelated no longer holds.  However the analysis can be
generalized to the colored case if we make assumptions that are motivated by
the properties of stationary detector noise.

In explaining this, it helps to begin by describing the stationary Gaussian
case.  For a colored Gaussian process, the Probability Density Function (PDF)
of the detector samples may be expressed as
\begin{equation}
  p(\vv{x}) = (2\pi)^{-N/2} (\det\vv{R})^{-1/2}
  \exp(-\frac{1}{2}\vv{x}^\dag\cdot\vv{Q}\cdot\vv{x})
\end{equation}
where the $N\times N$ correlation matrix
$\vv{R}=\langle\vv{x}\otimes\vv{x}^\dag\rangle$ is a positive-definite real
symmetric matrix with $N(N+1)/2$ real degrees of freedom and
$\vv{Q}=\vv{R}^{-1}$.  We have assumed that the process has zero mean.  The
volume element associated with this PDF is $d\vv{x}=\prod_{j=0}^{N-1}dx_j$.
In the time-domain, $\vv{x}$ is a vector of real numbers so
$\vv{x}^\dag=\vv{x}^\T$.

In the case where the random process is stationary, the matrix $\vv{R}$ is a
Toeplitz matrix, which depends only upon $|i-j|$.  Such a process is defined
by the first row or first column of the matrix and has only $N$ real degrees
of freedom.  Thus stationary Gaussian processes are a tiny subset of
\emph{all} Gaussian processes.

Now consider the PDF of new random variables that are linear combinations of
the old ones: $\tilde\vv{x}=\vv{U}\cdot\vv{x}$.  Take $\vv{U}$ to be an
arbitrary unitary matrix.  Clearly the PDF of these new variables
$\tilde\vv{x}$ is still Gaussian.  The matrix $\vv{U}$ can be chosen to
diagonalize the correlation matrix: this is called a Karhunen-Loeve
transformation.  In the limit where the time interval occupied by the $N$
samples is much larger than the correlation time of the noise, the linear
combinations of random variables that diagonalize the correlation matrix
asymptotically approach the Discrete Fourier Transform (DFT).  This is given
by
\begin{equation}
  U_{jk} = N^{-1/2}e^{2\pi ijk/N}.
\end{equation}
Thus, if $N$ is sufficiently large, to a good approximation the PDF of the new
variables in the Gaussian case may be written as
\begin{equation}
\label{e:fspacegauss}
  p(\tilde\vv{x}) = \prod_{k=1}^{[(N-1)/2]} 2\pi^{-1}P_k^{-2}
  \exp(-2|\tilde x_k|^2/P_k)
\end{equation}
where $P_k$ is the (real, positive) mean spectral amplitude in the $k$th
frequency bin:
\begin{equation}
\label{e:corrmaxf}
  \langle\tilde x_k \tilde x_{k'}^\ast\rangle = \frac{1}{2}\delta_{kk'}P_k
\end{equation}
for $1\le k,k'\le[(N-1)/2]$; thus
$\tilde\vv{R}=\vv{U}\cdot\vv{R}\cdot\vv{U}^{-1}\simeq\frac{1}{2}
{\mathrm{diag}}[P_k]$.  In other words, it is a good approximation to express
the PDF of a stationary colored Gaussian process as a diagonal process in
frequency space.

The limits of the product in Eq.~(\ref{e:fspacegauss}) appear strange because
$\tilde\vv{x}$ can not take on arbitrary values since $\vv{x}$ is real.  The
consequences include:
\begin{itemize}
\item 
  $\tilde x_k = \tilde x_{N-k}^\ast$.  Hence the amplitudes of
  $\tilde x_k$ for $k=[N/2]+1,\ldots,N-1$ are completely determined
  by $\tilde x_k$ for $k=1,\ldots,[(N-1)/2]$.
\item
  $\tilde x_0$ and, for even $N$, $\tilde x_{N/2}$ are real.  However, we
  assume that the data set has had the mean value (DC term) removed:
  $\sum_{j=0}^{N-1}x_j=0$.  Since gravitational wave detectors are AC-coupled
  and have no useful low-frequency response, this is a valid assumption. It
  implies that $\tilde x_0$ is identically zero.  Second, when $N$ is even, we
  assume that there is no energy in the Nyquist frequency bin: $\tilde
  x_{N/2}$ also vanishes identically.  This is a very reasonable assumption,
  since an experiment will include an anti-aliasing filter whose response (as
  a function of frequency) falls off rapidly as the Nyquist frequency is
  approached.
\item
  The volume element associated with this PDF is therefore
  $\prod_{k=1}^{[(N-1)/2]}d(\Re\tilde x_k)d(\Im\tilde x_k)$.
\end{itemize}

The likelihood ratio in the case of colored, stationary, Gaussian noise is
\begin{eqnarray}
  \ln\Lambda&=&\ln p(\vv{x}-\epsilon\vv{s})-\ln p(\vv{x}) \nonumber\\
  &=& \epsilon \vv{s}^\dag\cdot\vv{Q}\cdot\vv{x}
  - \frac{1}{2}\epsilon^2\vv{s}^\dag\cdot\vv{Q}\cdot\vv{s}
\end{eqnarray}
or, in the frequency domain,
\begin{equation}
  \ln\Lambda=\text{(constant)}
  +\epsilon4\Re\sum_{k=1}^{[(N-1)/2]}\tilde s_k^\ast \tilde x_k/P_k.
\end{equation}
Thus, the matched filter statistic, with a weighting equal to the inverse of
the noise spectrum, is the optimal detection statistic.

This motivates a more general model for the statistical distribution of
colored non-Gaussian detector noise, assuming that it is still stationary.  In
this case, to good approximation, the two-point correlation matrix
$\langle\tilde x_k \tilde x_{k'}^\ast\rangle$ is diagonal.  There may be
higher-order correlations present between the Fourier amplitudes at different
frequencies, but we will assume that this additional correlation is
negligible, and that to a reasonable approximation the probability
distribution of the noise in the non-Gaussian case is described by a PDF in
which the different frequency components are independent:
\begin{equation}
\label{e:assumedform}
  p(\tilde\vv{x}) = \prod_{k=1}^{[(N-1)/2]} 2 \pi^{-1} P_k^{-2}
  \exp[-2g_k(|\tilde x_k|^2/P_k)],
\end{equation}
with volume element $\prod_{k=1}^{[(N-1)/2]}d(\Re\tilde x_k)d(\Im\tilde x_k)$.
The functions $g_k(u)$ depend upon the frequency bin index $k$, so that the
statistical distribution can depend upon the frequency.  For the colored
Gaussian case the functions are $g_k(u)=u$.  In order that the PDF be properly
normalized, and that
$\langle\tilde x_k \tilde x_{k'}^\ast\rangle=\frac{1}{2}\delta_{kk'}P_k$, the
functions $g_k(u)$ must obey
\begin{equation}
\label{gnorm}
  \int_0^\infty e^{-g_k(u)}du =
  \int_0^\infty ue^{-g_k(u)}du = 1.
\end{equation}
Respectively, these constrain the additive constant in the definition of
$g_k$, and the multiplicative scale of the argument of $g_k$.  This is
\emph{not} the most general possible form of the probability distribution of a
stationary random process, but in many situations it should be a reasonable
approximation, particularly if the quantities of interest are dominated by the
second moments.

The locally optimal statistic may now be easily derived.  Letting $\tilde s_k$
denote the DFT of the expected waveform, and as before zeroing its DC and
Nyquist components, the conditional probability distribution of the detector
output is given by
\[
  p(\tilde\vv{x}|\epsilon) = \prod_{k=1}^{[(N-1)/2]} 2 \pi^{-1} P_k^{-2} 
  \exp[-2g_k(|\tilde x_k-\epsilon\tilde s_k|^2/P_k)]
\]
The locally optimal test can then be obtained from the first derivative:
\begin{equation}
  \Lambda_{(1)} = 4\sum_{k=1}^{[(N-1)/2]}\Re(\tilde s_k^\ast \tilde x_k/P_k)
    g_k'(|\tilde x_k|^2/P_k).
\end{equation}
In the colored Gaussian case $g_k'(u)=1$ this is the ordinary optimal linear
matched filter.  The contributions of the different frequency bins are
weighted by the inverse noise power spectrum in that bin.  In the non-Gaussian
case, just as for the case of uncolored white noise, the correlation in
frequency space is clipped or truncated for (frequency-bin) samples that lie
outside the central Gaussian part of the probability distribution, where
$|g'(u)|\ll1$.  An example of this may be seen in Fig.~\ref{f:weight2}: for
the illustrated case $g'(x^2/2) = f'(x)/x$.

Let us consider another form of non-Gaussian noise that describes a process in
which there is an ambient Gaussian noise background interrupted occasionally
by a large noise burst, which we will model as a second component of Gaussian
noise with a much larger variance.  The probability distribution we adopt
is~\cite{creighton99}:
\begin{eqnarray}
  p(\vv{x}) &=& (1-P)(2\pi)^{-{N \over 2}}(\det\vv{R})^{-{1 \over 2}}
  \exp(-\frac{1}{2}\vv{x}^\dag\cdot\vv{Q}\cdot\vv{x}) \nonumber \\
  &&+ P(2\pi)^{-{N \over 2}}(\det\bar\vv{R})^{-{1 \over 2}}
  \exp(-\frac{1}{2}\vv{x}^\dag\cdot\bar\vv{Q}\cdot\vv{x})
\end{eqnarray}
where $\vv{R}$ is the autocorrelation matrix for the normal ambient detector
noise and $\bar\vv{R}$ is the composite autocorrelation matrix for the
detector noise when a noise burst is present.  The noise bursts occur with
probability $P$ in this model.  Also, $\vv{Q}=\vv{R}^{-1}$ and
$\bar\vv{Q}=\bar\vv{R}^{-1}$.  We assume that $\bar\vv{Q}$ is much smaller
than $\vv{Q}$, meaning that
$\vv{x}^\dag\cdot\vv{Q}\cdot\vv{x}\gg\vv{x}^\dag\cdot\bar\vv{Q}\cdot\vv{x}$
for all vectors $\vv{x}$.  The locally optimal statistic is
\begin{equation}
  \Lambda_{(1)} =
  \left.\frac{d\ln p(\vv{x}-\epsilon\vv{s})}{d\epsilon}\right|_{\epsilon=0}
  = \frac{\Re(\vv{s}^\dag\cdot\vv{Q}\cdot\vv{x})}{1+\alpha}
\end{equation}
where
\begin{equation}
  \alpha = \frac{P}{1-P}\sqrt{\frac{\det\vv{R}}{\det\bar\vv{R}}}
  \exp[\frac{1}{2}\vv{x}^\dag\cdot(\vv{Q}-\bar\vv{Q})\cdot\vv{x}]
\end{equation}
is a detector of possible bursts.  When a burst is absent, $\alpha$ is
typically small and the locally optimal statistic reduces to the matched
filter.  However, when a burst is present, $\alpha$ is typically large and the
matched filter is suppressed.  Thus the locally optimal statistic is nearly
equivalent to the matched filter statistic with a veto if a segment of data
has a large amount of excess power as measured by
\begin{equation}
  {\mathcal{E}} = \vv{x}^\dag\cdot\vv{Q}\cdot\vv{x}
\end{equation}
or, in the frequency domain,
\begin{equation}
  {\mathcal{E}} = 4\sum_{k=1}^{[(N-1)/2]} |\tilde x_k|^2/P_k.
\end{equation}
The lengths (in time) of the data chunks used to estimate the
autocorrelation matrices should be choosen to be significantly longer
than the characteristic time of the signals being searched for, but
still short enough that the detector behavior is quasi-stationary.
For inspiral signals, typical signals are in the detector band for
tens of seconds, so the matrix estimation time should be at least of
order tens of minutes.  For stochastic background detection, the
correlation time between the two instruments is tens of milliseconds,
so that the matrix estimation time should be at least a few seconds.

Based on the two forms of non-Gaussian noise considered in this section, it
seems reasonable to adopt the following detection rules: (i) veto immediately
any segment of data that has an excess of power as measured by the excess
power statistic; (ii) for segments of data without an excess of power,
construct the matched filter in the frequency domain, but exclude those
frequency bins in which the detector power is too large.  The resulting
(truncated) matched filter is a good approximation of the locally optimal
statistic for a wide variety of possible non-Gaussian noise distribution.  In
this sense, it is a robust, nearly optimal detection statistic.

\section{STOCHASTIC SIGNALS}
\label{s:stochastic}
Observational limits from nucleosynthesis demonstrate that the stochastic
background of gravitational radiation has such small amplitude that it would
not be detectable with a single instrument \cite{allen96}.  In a single
instrument, there would be no practical way to discriminate between intrinsic
detector noise and the small additional noise-like output arising from a
stochastic background.  However, one can correlate the outputs of two
different instruments and search for a common signal in this way.  If the
instrumental noise is not Gaussian, then the previous single-detector analysis
can be easily generalized.

\subsection{Two coincident co-aligned detectors, white noise}
\label{ss:2dwnss}
We begin by considering the simple case in which the two detectors are
coincident and co-aligned, so that they have identical output contributions
from the stochastic background but independent intrinsic noise. We also assume
that the intrinsic noise samples in each detector are independent, and hence
white.

If the signal were deterministic (known) then the joint probability
distribution for the samples in the two detectors could be written as
\begin{equation}
\label{e:simplecase}
  p(\vv{x}_1,\vv{x}_2|\epsilon) = \prod_{i=0}^{N-1}
  e^{-f_1(x_{1,i}-\epsilon s_i)} e^{-f_2(x_{2,i}-\epsilon s_i)}
\end{equation}
This system can be analyzed in exactly the same way as in
Sec.~\ref{ss:sdwnkw}.  However the stochastic background does not produce a
known (deterministic) signal, so that the probability distribution needs to be
averaged over its expected distribution $p_{\text{sb}}(s_0,\ldots,s_{N-1})$
(which, by reason of the central limit theorem, is almost certainly a
multivariate Gaussian).  This leads to a joint probability distribution which
is given by
\begin{eqnarray}
\label{e:hardercase}
  p(\vv{x}_1,\vv{x}_2|\epsilon) &=& 
  \int ds_0 \cdots \int ds_{N-1} p_{\text{sb}}(s_0,\ldots,s_{N-1}) \nonumber\\
  & & \times\prod_{i=0}^{N-1}
  e^{-f_1(x_{1,i}-\epsilon s_i)-f_2(x_{2,i}-\epsilon s_i)} \\
  & = & \int d\vv{s} p_{\text{sb}}(\vv{s}) \prod_{i=0}^{N-1}
  e^{-f_1(x_{1,i}-\epsilon s_i)-f_2(x_{2,i}-\epsilon s_i)} \nonumber
\end{eqnarray}
Here $\epsilon$ may be thought of as the coupling of the detector.  The case
of small $\epsilon$ corresponds to a detector that is only weakly coupled to
the signal. For this non-deterministic signal, it is still straightforward to
construct a locally optimal test, and a corresponding decision statistic or
threshold criterion.

The locally optimal statistic is obtained from the derivative of the
probability distribution with respect to $\epsilon$.  This is given by
\begin{eqnarray}
\label{e:firstder}
  \frac{dp(\vv{x}_1,\vv{x}_2|\epsilon)}{d\epsilon} &&=
  \int d\vv{s} p_{\text{sb}}(\vv{s}) \nonumber\\
  & & \times\left( \sum_{j=0}^{N-1}
  s_j [ f_1'(x_{1,j}-\epsilon s_j) + f_2'(x_{2,j}-\epsilon s_j])  \right)
  \nonumber\\
  & & \times\prod_{i=0}^{N-1} 
  e^{-f_1(x_{1,i}-\epsilon s_i)-f_2(x_{2,i}-\epsilon s_i)}.
\end{eqnarray}
Setting $\epsilon=0$ and dividing by $p(\vv{x}_1,\vv{x}_2|0)$ yields the
locally optimal statistic:
\begin{equation}
  \Lambda_{(1)}=\sum_{j=0}^{N-1} [f_1'(x_{1,j}) + f_2'(x_{2,j})]
  \times\int s_j p_{\text{sb}}(\vv{s}) d\vv{s}.
\end{equation}
Unfortunately this vanishes if the random process described by
$p_{\text{sb}}(\vv{s})$ has vanishing mean, since in this case
$\int s_jp_{\text{sb}}(\vv{s})d\vv{s}=0$.  This is indeed the case for the
gravitational-wave stochastic background.

When the \emph{first} derivative vanishes, the locally optimal statistic is
defined by having the largest \emph{second} derivative at $\epsilon=0$.  See
Fig.~\ref{f:powerfn} for example. Taking another derivative of
(\ref{e:firstder}) and setting $\epsilon=0$ yields
\begin{equation}
  \Lambda_{(2)} = \int d\vv{s} p_{\text{sb}}(\vv{s})
  \begin{array}[t]{l}
    \biggl\{\Bigl(\sum_{j=0}^{N-1} s_j [f_1'(x_{1,j})+f_2'(x_{2,j})] \Bigr)^2\\
    \quad-\sum_{j=0}^{N-1}s_j^2[f_1''(x_{1,j})+f_2''(x_{2,j})] \biggr\}
  \end{array}
\end{equation}
The terms that appear in this statistic have different character, and before
moving on, some discussion is required.

The locally optimal statistic depends upon the statistical character of the
stochastic background radiation through the second-order moments.  We will
assume that the stochastic background is a stationary process, so that the
second order correlation $\langle s_i s_j\rangle$ is a function of the lag
$|i-j|$ only:
\begin{equation}
  C(|i-j|)=\langle s_i s_j \rangle = \int d\vv{s}p_{\text{sb}}(\vv{s})s_i s_j.
\end{equation}
In a stochastic background search, the ``signal model'' only requires an
assumption about the form of the spectrum.  This is (roughly) the Fourier
transform of $C(\Delta)$.  Without loss of generality we normalize $C(\Delta)$
so that $C(0)=1$ (this simply scales the value of $\epsilon$).  Expressing the
locally optimal statistic in terms of the correlation function $C$ then gives
\begin{eqnarray}
\label{e:lostochbg}
  \Lambda_{(2)}&=&-\sum_{i=0}^{N-1}[f_1''(x_{1,i})+f_2''(x_{2,i})] \nonumber\\
  && + \sum_{j,k=0}^{N-1} C(|j-k|) [ f_1'(x_{1,j}) f_1'(x_{1,k}) \\
  && \qquad + f_2'(x_{2,j}) f_2'(x_{2,k}) + 2 f_1'(x_{1,j}) f_2'(x_{2,k})].
\nonumber
\end{eqnarray}
Each of the five terms that appears in (\ref{e:lostochbg}) has a specific
interpretation. The first four terms that appear in the locally optimal
estimator $\Lambda_{(2)}$ are generalized ``single-detector'' statistics which
do not cross-correlate the two detectors.  They are generalized measures of
the ``energy'' received by each individual detector, and provide useful
information only if the stochastic background contributes substantially more
to the measured signal than the detector output does, or if the detector's
intrinsic noise contributions can somehow be separated from the noise
contribution arising from the stochastic background.  (This will not be the
case for the first few generations of gravitational wave detectors).  The last
term in (\ref{e:lostochbg}) is a Generalized Cross-Correlation (GCC)
statistic, that provides useful information even if the detector noise
dominates the signal: the expected case for gravitational-wave stochastic
background.  To quote from Kassam (following Eq.~(7-24) in Ref.~\cite{kassam})
\begin{quotation}
\noindent
It is important to note that the increase in power level occurs whenever
random signals are present at the individual receivers of the array regardless
of whether the signals \emph{across} the array are one common signal or are
completely uncorrelated.  The GCC part of the Locally Optimal (LO) statistic
responds only to a \emph{common} signal or at least to signals which are
spatially correlated across the array elements.  This is a major reason why it
is useful to employ only the GCC part of the LO statistic in applications
involving detection as well as location of signal sources.
\end{quotation}
For this reason
(and others \footnote{
There are other motivations for dropping the remaining terms.  In particular,
the remaining terms are non-zero in the absence of correlated stochastic
background: the GCC term vanishes in this case.  One can show that terms 2, 3,
and 4 are all positive definite (since $\langle s_i s_j \rangle$ is positive
definite) and that for the expected ``Gaussian + tail'' detector probability
distribution, term 1 is also positive definite.
Baysian analysis (in part II of this series of papers) also
shows that the single-detector terms must be dropped.  This may also
be seen in the analysis at the end of Sec.~\ref{ss:estimators}.})
we drop the single-detector
terms from the statistic,
and define the GCC statistic as:
\begin{equation}
\label{e:gcc}
  \Lambda_{\text{GCC}}= 2\sum_{j,k=0}^{N-1}C(|j-k|)f_1'(x_{1,j})f_2'(x_{2,k}).
\end{equation}
This generalized cross-correlation statistic reduces to the ordinary
cross-correlation statistic in the case where the detector noise is Gaussian:
$f_1(x)=f_2(x)=x^2/2+\log(2\pi)/2$.  It can be easily generalized to the case
of three or more detectors \cite{kassam}.

In practical work $C$ will vanish for lags greater than the light travel time
between the two detectors (i.e., 10~ms for the LIGO detectors).  This means
that even if $N$ is chosen to be very large, $\Lambda_{\text{GCC}}$ only
correlates samples from the two detectors taken within this time window.
(Note: if the detector noise is colored, then the time window may be larger,
as will be seen shortly.)

\subsection{Two non-coincident non-co-aligned detectors, colored noise}
\label{ss:2dcnss}
In this section, we generalize the work of Sec.~\ref{ss:2dwnss} to the case
where the two detectors are not coincident or co-aligned, and their noise
power spectrum is not white. We assume that the intrinsic detector noise of
the two detectors is independent.  If the two detectors are widely separated
and subject to different environmental influences, this assumption should
hold.

Let us start by assuming that the two detectors each have internal
(instrumental) colored Gaussian noise with known autocorrelation matrices
$\vv{R}_{\text{in},1}$ and $\vv{R}_{\text{in},2}$, and that the instrumental
noise of the two detectors is independent.  The stochastic background produces
an additional source of colored Gaussian noise that is correlated between the
two detectors.  The stochastic background noise is measured by the
autocorrelation matrices
$\vv{S}_{11}=\langle\vv{s}_1\otimes\vv{s}_1^\dag\rangle$,
$\vv{S}_{22}=\langle\vv{s}_2\otimes\vv{s}_2^\dag\rangle$, and the
cross-correlation matrices
$\vv{S}_{12}=\langle\vv{s}_1\otimes\vv{s}_2^\dag\rangle$,
$\vv{S}_{21}=\langle\vv{s}_2\otimes\vv{s}_1^\dag\rangle$.  Since the stochastic
background is isotropic, $\vv{S}_{11}=\vv{S}_{22}=\vv{R}_{\text{sb}}$ (the
stochastic background contribution to the detector's autocorrelation matrices)
and $\vv{S}_{12}=\vv{S}_{21}=\vv{S}$ (the cross-correlated noise between the
detectors due to the stochastic background).  The total autocorrelation noise
of the two detectors are
$\vv{R}_1=\vv{R}_{\text{in},1}+\epsilon^2\vv{R}_{\text{sb}}$ and
$\vv{R}_2=\vv{R}_{\text{in},2}+\epsilon^2\vv{R}_{\text{sb}}$.  In the presence
of the stochastic background, the likelihood ratio is
\begin{equation}
  p(\vv{x}_1,\vv{x}_2|\epsilon) = (2\pi)^{-N}(\det\vv{\Sigma})^{-1}
  \exp(-\frac{1}{2}\bs{\xi}^\dag\cdot\vv{\Sigma}^{-1}\cdot\bs{\xi})
\end{equation}
where
\begin{displaymath}
  \bs{\xi} = \left[\begin{array}{c}\vv{x}_1\\\vv{x}_2\end{array}\right]
  \quad\text{and}\quad
  \vv{\Sigma} = \left[
  \begin{array}{cc}
    \vv{R}_1 & \epsilon^2\vv{S} \\ \epsilon^2\vv{S} & \vv{R}_2
  \end{array} \right].
\end{displaymath}
In the weak signal approximation,
\begin{eqnarray}
  \vv{\Sigma}^{-1} &=& \left[
  \begin{array}{cc}
    \vv{Q}_{\text{in},1} & \vv{0} \\ \vv{0} & \vv{Q}_{\text{in},2}
  \end{array} \right] \nonumber\\
  &&- \epsilon^2 \left[
  \begin{array}{cc}
    \vv{Q}_{\text{in},1}\cdot\vv{R}_{\text{sb}}\cdot\vv{Q}_{\text{in},1} &
    \vv{Q}_{\text{in},1}\cdot\vv{S}\cdot\vv{Q}_{\text{in},2}             \\
    \vv{Q}_{\text{in},2}\cdot\vv{S}\cdot\vv{Q}_{\text{in},1}             &
    \vv{Q}_{\text{in},2}\cdot\vv{R}_{\text{sb}}\cdot\vv{Q}_{\text{in},2}
  \end{array} \right] \nonumber\\
  && + \epsilon^4 \biggl\{ \left[
  \begin{array}{c}
      \vv{Q}_{\text{in},1}\cdot\vv{R}_{\text{sb}}\cdot\vv{Q}_{\text{in},1}
      \cdot\vv{R}_{\text{sb}}\cdot\vv{Q}_{\text{in},1} ,\quad \vv{0} \\
      \vv{0} ,\quad \vv{Q}_{\text{in},2}\cdot\vv{R}_{\text{sb}}\cdot
      \vv{Q}_{\text{in},2}\cdot\vv{R}_{\text{sb}}\cdot\vv{Q}_{\text{in},2}
  \end{array}\right]
  \nonumber\\
  &&\quad + \left[
  \begin{array}{c}
      \vv{Q}_{\text{in},1}\cdot\vv{S}\cdot
      \vv{Q}_{\text{in},2}\cdot\vv{S}\cdot\vv{Q}_{\text{in},1} ,\quad \vv{0} \\
      \vv{0} ,\quad \vv{Q}_{\text{in},2}\cdot\vv{S}\cdot
      \vv{Q}_{\text{in},1}\cdot\vv{S}\cdot\vv{Q}_{\text{in},2}
  \end{array} \right] \biggr\} \nonumber\\
  && + O(\epsilon^6),
\end{eqnarray}
\begin{eqnarray}
  \ln\det\vv{\Sigma}&=&\ln\det\vv{R}_{\text{in},1}+\ln\det\vv{R}_{\text{in},2}
  \nonumber\\
  && + \epsilon^2[\tr(\vv{Q}_{\text{in},1}\cdot\vv{R}_{\text{sb}})
     + \tr(\vv{Q}_{\text{in},2}\cdot\vv{R}_{\text{sb}})] \nonumber\\
  && - \epsilon^4[
       \frac{1}{2}\tr(\vv{Q}_{\text{in},1}\cdot\vv{R}_{\text{sb}}
           \cdot\vv{Q}_{\text{in},1}\cdot\vv{R}_{\text{sb}}) \nonumber\\
  &&\qquad + \frac{1}{2}\tr(\vv{Q}_{\text{in},2}\cdot\vv{R}_{\text{sb}}
           \cdot\vv{Q}_{\text{in},2}\cdot\vv{R}_{\text{sb}}) \nonumber\\
  &&\qquad + \tr(\vv{Q}_{\text{in},2}\cdot\vv{S}\cdot\vv{Q}_{\text{in},1}
           \cdot\vv{S})] \nonumber\\
  && + O(\epsilon^6),
\end{eqnarray}
and
\begin{eqnarray}
\ln\Lambda &=& \ln p(\vv{x}_1,\vv{x}_2|\epsilon) - \ln p(\vv{x}_1,\vv{x}_2|0)
\nonumber\\
&=& \epsilon^2\{ -\frac{1}{2}\tr(\vv{Q}_{\text{in},1}\cdot\vv{R}_{\text{sb}})
   -\frac{1}{2}\tr(\vv{Q}_{\text{in},2}\cdot\vv{R}_{\text{sb}}) \nonumber\\
&&\qquad + \Re(\vv{x}_2^\dag\cdot\vv{Q}_{\text{in},2}\cdot\vv{S}
    \cdot\vv{Q}_{\text{in},1}\cdot\vv{x}_1) \nonumber\\
&&\qquad +\frac{1}{2}\vv{x}_1^\dag\cdot\vv{Q}_{\text{in},1}
    \cdot\vv{R}_{\text{sb}}\cdot\vv{Q}_{\text{in},1}\cdot\vv{x}_1 \\
&&\qquad +\frac{1}{2}\vv{x}_2^\dag\cdot\vv{Q}_{\text{in},2}
    \cdot\vv{R}_{\text{sb}}\cdot\vv{Q}_{\text{in},2}\cdot\vv{x}_2 \}
    + O(\epsilon^4) \nonumber
\end{eqnarray}
where $\vv{Q}_{\text{in},1}=\vv{R}_{\text{in},1}^{-1}$
and $\vv{Q}_{\text{in},2}=\vv{R}_{\text{in},2}^{-1}$.  The last two terms
represent the autocorrelation ``energy'' detectors.

The following question now becomes important: how does one obtain the
quantities $\vv{R}_{\text{in},1}$ and $\vv{R}_{\text{in},2}$.  There are two
possible methods: (i) by a theoretical understanding of the detector, or (ii)
by shielding the instrument from the stochastic background and measuring the
noise autocorrelation.  For gravitational wave searches, method (ii) is not
available as there is no way to shield the detector from a stochastic
background of gravitational waves.  Method (i) holds more promise, but if the
stochastic background is expected to be weak, it is unlikely that our
understanding of the detector will be sufficient to distinguish between the
noise autocorrelations $\vv{R}_{\text{in}}$ and
$\vv{R}_{\text{in}}+\epsilon\vv{R}_{\text{sb}}$.  We expect that the noise
matrices that should be used are the \emph{measured} noise matrices
$\vv{R}_1=\langle\vv{x}_1\otimes\vv{x}_1^\dag\rangle$ and
$\vv{R}_2=\langle\vv{x}_2\otimes\vv{x}_2^\dag\rangle$, which contain both the
internal, instrumental noise as well as the stochastic background ``noise.''
Since it is these quantities rather than $\vv{R}_{\text{in},1}$ and
$\vv{R}_{\text{in},2}$ that are known, the previous analysis must be modified.
We now have
\begin{eqnarray}
  \vv{\Sigma}^{-1} &=& \left[
  \begin{array}{cc}
    \vv{Q}_1 & \vv{0} \\ \vv{0} & \vv{Q}_2
  \end{array} \right]
  - \epsilon^2 \left[
  \begin{array}{cc}
    \vv{0} & \vv{Q}_1\cdot\vv{S}\cdot\vv{Q}_2  \\
    \vv{Q}_2\cdot\vv{S}\cdot\vv{Q}_1 & \vv{0}
  \end{array} \right] \nonumber\\
  && + \epsilon^4 \left[
  \begin{array}{c}
      \vv{Q}_1\cdot\vv{S}\cdot\vv{Q}_2\cdot\vv{S}\cdot\vv{Q}_1,\quad \vv{0} \\
      \vv{0},\quad \vv{Q}_2\cdot\vv{S}\cdot\vv{Q}_1\cdot\vv{S}\cdot\vv{Q}_2
  \end{array} \right] \nonumber\\
  && + O(\epsilon^6),
\end{eqnarray}
\begin{eqnarray}
  \ln\det\vv{\Sigma} &=& \ln\det\vv{R}_1+\ln\det\vv{R}_2 \nonumber\\
  && - \epsilon^4 \tr(\vv{Q}_2\cdot\vv{S}\cdot\vv{Q}_1\cdot\vv{S})
  + O(\epsilon^6),
\end{eqnarray}
and
\begin{equation}
  \ln\Lambda = \epsilon^2
    \Re(\vv{x}_2^\dag\cdot\vv{Q}_2\cdot\vv{S}\cdot\vv{Q}_1\cdot\vv{x}_1)
    + O(\epsilon^4)
\end{equation}
where $\vv{Q}_1=\vv{R}_1^{-1}$ and $\vv{Q}_2=\vv{R}_2^{-1}$.  The locally
optimal detection statistic (which is appropriate for weak signals) is the
cross-correlation statistic.

\ifmgf 

To generalize to non-Gaussian noise, it is helpful to use moment generating
functions.  Suppose the vector $\vv{n}_1$ represents the internal
(instrumental) noise in the first detector.  The moment generating function
for $\vv{n}_1$ is
\begin{equation}
  \Phi_{\text{in},1}(\vv{w}_1)=\langle e^{i\vv{w}^\T\cdot\vv{n}_1} \rangle
\end{equation}
and the probability distribution for $\vv{n}_1$ is the Fourier transform of
the moment generating function:
\begin{equation}
  p_{\text{in},1}(\vv{n}_1)=\int d\vv{w}_1 e^{-i\vv{n}_1^\T\cdot\vv{w}}
    \Phi_{\text{in},1}.
\end{equation}
The moment generating function $\Phi_{\text{in},2}(\vv{w}_2)$ for the internal
noise in detector 2 is defined similarly.  We assume that the stochastic
background is a multivariate Gaussian with a moment generating function
\begin{equation}
  \Phi_{\text{sb}}(\vv{w}_1,\vv{w}_2)
  = \exp(-\frac{1}{2}\epsilon^2
  \bs{\omega}^\T\cdot\vv{\Sigma}_{\text{sb}}\cdot\bs{\omega})
\end{equation}
with
\begin{displaymath}
  \bs{\omega}=\left[\begin{array}{c}\vv{w}_1 \\ \vv{w}_2\end{array}\right]
  \quad\text{and}\quad
  \vv{\Sigma}_{\text{sb}}=\left[
  \begin{array}{cc}
    \vv{R}_{\text{sb}} & \vv{S} \\ \vv{S} & \vv{R}_{\text{sb}}
  \end{array} \right].
\end{displaymath}
Then the moment generating function for the detectors' output is
\begin{eqnarray}
  \Phi(\vv{w}_1,\vv{w}_2) &=&
  \langle e^{i\vv{w}_1^\T\cdot\vv{x}_1} e^{i\vv{w}_2^\T\cdot\vv{x}_2} \rangle
  \\
  &=& \Phi_{\text{in},1}(\vv{w}_1) \Phi_{\text{in},2}(\vv{w}_2)
  \Phi_{\text{sb}}(\vv{w}_1,\vv{w}_2) \nonumber
\end{eqnarray}
and the joint probability distribution is
\begin{eqnarray}
  p(\vv{x}_1,\vv{x}_2|\epsilon) &=&
  \int d\vv{w}_1 d\vv{w}_2
  e^{-i(\vv{x}_1^\T\cdot\vv{w}_1+\vv{x}_2^\T\cdot\vv{w}_2)}
  \Phi(\vv{w}_1,\vv{w}_2) \nonumber\\
  &=& \int d\bs{\omega} \exp(-i\bs{\xi}^\T\cdot\bs{\omega})
  \Phi_{\text{in},1}(\vv{w}_1) \Phi_{\text{in},2}(\vv{w}_2) \nonumber\\
  &&\times\{1 - \frac{1}{2}\epsilon^2
  \bs{\omega}^\T\cdot\vv{\Sigma}_{\text{sb}}\cdot\bs{\omega}
  + O(\epsilon^4) \} \nonumber\\
  &=& p(\vv{x}_1,\vv{x}_2|0) \nonumber\\
  &&+ \epsilon^2\{
  (\bs{\nabla}^\T p_{\text{in},2})(\vv{x}_2)\cdot\vv{S}
  \cdot(\bs{\nabla}p_{\text{in},1})(\vv{x}_1) \nonumber\\
  &&\quad+
  \frac{1}{2}(\bs{\nabla}^\T p_{\text{in},1})(\vv{x}_1)\cdot\vv{R}_{\text{sb}}
  \cdot(\bs{\nabla}p_{\text{in},1})(\vv{x}_1) \nonumber\\
  &&\quad+
  \frac{1}{2}(\bs{\nabla}^\T p_{\text{in},2})(\vv{x}_2)\cdot\vv{R}_{\text{sb}}
  \cdot(\bs{\nabla}p_{\text{in},2})(\vv{x}_2)\} \nonumber\\
  &&+O(\epsilon^4).
\end{eqnarray}
Thus, if we ignore the autocorrelation terms, the locally optimal statistic is
\begin{equation}
  \Lambda_{(2)}=
  (\bs{\nabla}^\T\ln p_{\text{in},2})(\vv{x}_2)\cdot\vv{S}
  \cdot(\bs{\nabla}\ln p_{\text{in},1})(\vv{x}_1).
\end{equation}
This equation for the locally optimal statistic is good for the time domain,
in which the detectors' output vectors are real and so the derivative is
meaningful.

To extend the result to complex vectors, and thus to a frequency-domain
representation, we use the following formal replacement: replace every complex
number $x=a+ib$ and derivative $\nabla$ with the matrices
\begin{displaymath}
  x \to \underline x = \left[\begin{array}{rr}a&b\\-b&a\end{array}\right]
  \quad\text{and}\quad
  \nabla \to \underline\nabla = \frac{1}{2} \left[
  \begin{array}{rr}
    \partial/\partial a & -\partial/\partial b \\
    \partial/\partial b & \partial/\partial a
  \end{array} \right].
\end{displaymath}
Note that this means $x^\ast$ is represented by $\underline x^\T$ and $|x|^2$
by $\underline x^\T\cdot\underline x$.  Also, the meaning of $\nabla|x|^2$ is
$\underline\nabla(\underline x^\T\cdot\underline x)=\underline x^\T$.  The
locally optimal statistic is
\begin{eqnarray}
  \Lambda_{(2)} = \frac{1}{2}
  (\underline{\bs{\nabla}}^\T\ln p_{\text{in},2})(\underline{\vv{x}}_2)
  \cdot&\underline{\vv{S}}&
  \cdot(\underline{\bs{\nabla}}\ln p_{\text{in},1})(\underline{\vv{x}}_1)
  \nonumber\\
  + \frac{1}{2}
  (\underline{\bs{\nabla}}\ln p_{\text{in},2})(\underline{\vv{x}}_2)\cdot
  &\underline{\vv{S}}&
  \cdot(\underline{\bs{\nabla}}^\T\ln p_{\text{in},1})(\underline{\vv{x}}_1).
\end{eqnarray}
For example, for the noise model in which 
$\ln p_{\text{in}}(\tilde\vv{x})\propto
\sum_{k=1}^{[(N-1)/2]}g_k(|\tilde x_k|^2/2P_k)$ and
$\tilde\vv{S}={\mathrm{diag}}[\gamma_k\sigma_k^2]$, the locally optimal
statistic is
\begin{eqnarray}
  \Lambda_{(2)} &=& \Re\sum_{k=1}^{[(N-1)/2]}
  \frac{\gamma_k\sigma^2_k\tilde x_{1,k}^\ast\tilde x_{2,k}}{P_{1,k}P_{2,k}}
  \nonumber\\
  &&\quad\times
  g_{1,k}'(|\tilde x_{1,k}|^2/P_{1,k})
  g_{2,k}'(|\tilde x_{2,k}|^2/P_{2,k}).
\end{eqnarray}

\fi 

Before we examine specific non-Gaussian noise models, we will describe the
form of the matrices $\vv{R}_{\text{sb}}$ and $\vv{S}$.  A stochastic
background, if present, contributes to the signal amplitude at each detector.
To simplify the analysis, in Sec.~\ref{ss:2dwnss}, we assumed that the
detectors were coincident and co-aligned, so that the amplitude contribution
in each individual detector are identical.  Here, we drop that assumption.

Because the detectors are not co-aligned, the axes of the two interferometer
arms point in different directions, and are sensitive to different linear
combinations of the two possible gravitational wave polarizations.  This
reduces the correlation between the amplitudes in the two detectors, since we
will assume that the stochastic background is unpolarized. An additional loss
of correlation occurs because the two detectors are separated.  This loss of
correlation becomes increasingly greater for shorter wavelengths.  Roughly
speaking, there is no significant loss of correlation for wavelengths much
longer than the inter-detector distance, and there is a complete loss of
correlations for wavelengths much shorter than this\cite{allen96}.

The loss of amplitude correlation due to the separation and non-alignment of
the two detectors may be described (for an unpolarized and isotropic
stochastic background) in terms of the overlap reduction function $\gamma(f)$
defined by Flanagan \cite{flanagan93}. This quantity is the average value of
the product of the detector outputs, for a stochastic background of a given
frequency $f$, averaged over the possible directions of arrival and phases. It
is given by:
\begin{equation}
\label{e:overlap}
  \gamma(f) = \frac{5}{8\pi} \int_{S^2} d \hat \Omega \>
  e^{2 \pi i f \hat \Omega \cdot \Delta \vec x/c }
  ( F_1^+ F_2^+ + F_1^\times F_2^\times ).
\end{equation}
Here $\hat\Omega$ is a unit-length vector on the two-sphere, $\Delta\vec x$ is
the separation between the two detector sites, and $F_i^{+,\times}$ is the
response of detector $i$ to the $+$ or $\times$ polarization.  For the $i$th
detector $(i=1,2)$ one has
\begin{equation}
\label{e:detectresponse}
  F_i^{+,\times}
   = \frac{1}{2} ( \hat X_i^a \hat X_i^b - \hat Y_i^a \hat Y_i^b )
  e_{ab}^{+,\times}(\hat \Omega),
\end{equation}
where $e_{ab}^{+,\times}(\hat\Omega)$ are the gravitational wave polarization
tensors for a wave propagating in direction $\hat\Omega$.  The normalization
of $\gamma(f)$ is chosen so that for coincident and co-aligned detectors,
$\gamma(f)=1$.  For co-aligned but not coincident detectors, $\gamma(f=0)=1$.
For coincident but unaligned detectors, $\gamma(f)$ is a frequency-independent
constant that depends only upon the relative orientation of the two detectors,
and vanishes if the two detectors are sensitive to orthogonal polarizations.

\begin{figure}
\epsfig{file=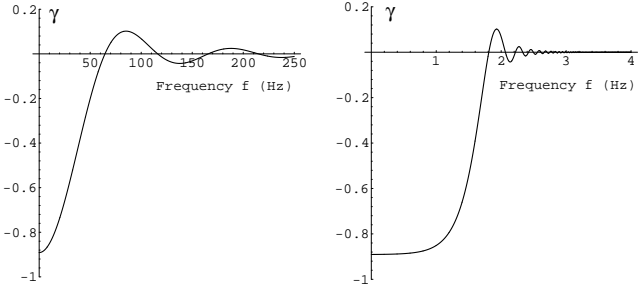,width=3.35in}
\caption{
\label{f:gammaligo}
The overlap reduction function $\gamma(f)$ is shown for the two LIGO detectors
as a function of frequency $f$. The left/right graphs have linear/$\log_{10}$
frequency axes. Because the detectors are almost anti-aligned, the function is
close to $-1$ at low frequencies.  The first root is at 64 Hz.}
\end{figure}

General expressions for $\gamma(f)$ for arbitrary detectors may be found in
Refs.~\cite{flanagan93,allen97}.  For the pair of LIGO detectors $\gamma(f)$
is shown in Fig.~\ref{f:gammaligo}, and is given by
\begin{equation}
  \gamma(f) \approx -0.1248 \> j_{0}(x) - 2.900 \> {{j_{1}(x)}\over x} + 
    3.008 \> {{j_{2}(x)}\over {{x^2}}}
\end{equation}
where $x=2\pi fd/c$ is a frequency variable, $d=3010\,\text{km}$ is the
detector separation, $c=2.998\times10^5\,\text{km/s}$ is the speed of light,
and $j_n$ is a spherical Bessel function.  It is helpful to introduce notation
for the overlap reduction function's values in the frequency bins of interest.
Let $f_k=k/(N\Delta t)=k/T$ denote the frequency of the $k$th bin, with
$k=0,\ldots,[N/2]$.  Here $\Delta t$ is the sample interval and $T=N\Delta t$
is the total observation time. Then
\begin{equation}
  \gamma_k = \gamma(f_k) = \gamma(k/T).
\end{equation}
are the values of the overlap reduction function in the $k$th bin.

The stochastic background is characterized by its dimensionless energy density
\begin{equation}
  \Omega_{\text{gw}}(f) = \frac{1}{\rho_{\text{critical}}}
  \frac{d\rho_{\text{gw}}}{d\ln f},
\label{e:Omega_gw}
\end{equation}
where $d\rho_{\text{gw}}$ is the energy density of the gravitational radiation
contained in the frequency range $f$ to $f+df$, and $\rho_{\text{critical}}$
is the critical energy density required (today) to close the universe:
\begin{equation}
  \rho_{\text{critical}} = \frac{3c^2H_0^2}{8\pi G} \approx 1.6\times
  10^{-8}h_{100}^2\,\text{erg}\,\text{cm}^{-3}.
\label{e:rho_crit}
\end{equation}
$H_0$ is the Hubble expansion rate (today):
\begin{equation}
  H_0 = h_{100} \times 100\,\text{km}\,\text{s}^{-1}\,\text{Mpc}^{-1}
  = 3.2 \times 10^{-18}h_{100}\,\text{s}^{-1},
\label{e:H_0}
\end{equation}
and $h_{100}$ is a dimensionless factor that we have included to account for
the different values of $H_0$ that are quoted in the literature.%
\footnote{$h_{100}$ is probably very close to 0.65.}

The PDF of the stochastic background strain can usually be expressed in closed
form.  The central limit theorem shows that if the stochastic background has
been produced (as it is in many scenarios) by an incoherent sum of many small
processes, then its statistics will be stationary\cite{allenflanaganpapa} and
Gaussian.  This means that it is characterized by the single-site second
moments
\begin{equation}
  \langle\tilde s_{1,k} \tilde s_{1,k'}^\ast\rangle
  = \langle\tilde s_{2,k} \tilde s_{2,k'}^\ast\rangle
  = \sigma^2_k \delta_{kk'},
\end{equation}
with
\begin{equation}
  \sigma^2_k=\frac{3H_0^2\Omega_{\text{gw}}(f_k)}{20\pi^2\Delta t|f_k|^3}.
\end{equation}
As before, we have assumed that $N$ is chosen so that $N\Delta t$ is much
larger than the correlation time of the stochastic background (filtered by the
instrument response function), so that the RHS of (\ref{e:sbg2nd}) is
proportional to $\delta_{kk'}$.  The expectation value of the product of the
strain at the two different sites is reduced by the overlap reduction
function:
\begin{equation}
\label{e:sbg2nd}
  \langle\tilde s_{1,k} \tilde s_{2,k'}^\ast\rangle
  = \langle\tilde s_{2,k} \tilde s_{1,k'}^\ast\rangle
  = \gamma_k \langle\tilde s_{1,k} \tilde s_{1,k'}^\ast\rangle
  = \gamma_k \sigma^2_k \delta_{kk'}.
\end{equation}
This follows from Eqn.~(3.56) of reference \cite{allen97}.  In practice, since
the shape of the stochastic background spectrum is not know, the dependence of
the $\sigma_k$ on $k$ should be assumed to fit some simple parameterized model,
such as a power law $\sigma^2_k \propto k^\alpha$ for a reasonable range of
$\alpha$.

We can now express the locally optimal detection statistic for a stochastic
background in colored Gaussian noise.  It is:
\begin{equation}
  \ln\Lambda = \epsilon^2\Re\sum_{k=1}^{[(N-1)/2]}
    \gamma_k \sigma^2_k x_{1,k}^\ast x_{2,k}/(P_{1,k}P_{2,k})
\end{equation}
where $P_{1,k}$ and $P_{2,k}$ are the measured noise spectra in the two
detectors.

Let us now turn to our first non-Gaussian noise model.  Our starting point is
a PDF for the noise in the two detectors in the absence of any stochastic
background signal.  We make the same assumptions about the detector noise as
in Sec.~\ref{ss:sdcnkw}.  The PDF is given in frequency space by a product of
two terms identical to (\ref{e:assumedform}),
\begin{eqnarray*}
  p(\tilde\vv{x}_1,\tilde\vv{x}_2) &=& \prod_{k=1}^{[(N-1)/2]}
    2 \pi^{-1} P_{1,k}^{-2} e^{-2g_{1,k}(|\tilde x_{1,k}|^2/P_{1,k})} \\
  && \times \prod_{k'=1}^{[(N-1)/2]}
    2 \pi^{-1} P_{2,k'}^{-2} e^{-2g_{2,k'}(|\tilde x_{2,k'}|^2/P_{2,k'})} \\
  &=& \prod_{k=1}^{[(N-1)/2]} 4 \pi^{-2} P_{1,k}^{-2} P_{2,k}^{-2} \\
  && \times e^{-2g_{1,k}(|\tilde x_{1,k}|^2/P_{1,k})
    -2g_{2,k}(|\tilde x_{2,k}|^2/P_{2,k})}.
\end{eqnarray*}

The statistical distribution of the stochastic background is
\begin{eqnarray}
  && p_{\text{sb}}(\tilde\vv{s}_1,\tilde\vv{s}_2) = 
  \prod_{k=1}^{[(N-1)/2]} (\pi\sigma_k^2)^{-2}(1-\gamma_k^2)^{-1} \nonumber\\
  && \quad \times \exp\left( - 
    \frac{|\tilde s_{1,k}|^2 + |\tilde s_{2,k}|^2
          - 2\gamma_k \Re(\tilde s_{1,k}^\ast \tilde s_{2,k})}%
         {\sigma_k^2 (1-\gamma_k^2)}
  \right).
\end{eqnarray}

We can now find the locally optimal statistic. Since the detector is linear,
as before, one has a joint probability distribution for the observed Fourier
amplitudes:
\begin{displaymath}
  p(\tilde\vv{x}_1,\tilde\vv{x}_2|\epsilon) =
  \int d\tilde\vv{s}_1 d\tilde\vv{s}_2
  p_{\text{sb}}(\tilde\vv{s}_1,\tilde\vv{s}_2)
  p(\tilde\vv{x}_1-\epsilon\tilde\vv{s}_1,
  \tilde\vv{x}_2-\epsilon\tilde\vv{s}_2).
\end{displaymath}
This corresponds to a stochastic background with a characteristic
energy-density function $\epsilon\Omega_{\text{gw}}(f)$.%
\footnote{In fact $\epsilon$ should really be interpreted as ``coupling to the
detector''.  The locally optimal statistic is the best choice in the ``weakly
coupled detector'' limit.}
The locally optimal statistic is
\begin{eqnarray} 
  \Lambda_{(1)} = 4\Re \sum_{k=1}^{[(N-1)/2]} &&\biggl\{
  \frac{\langle\tilde s_{1,k}^\ast\rangle \tilde x_{1,k}
        g_{1,k}'(|\tilde x_{1,k}|^2/P_{1,k})}{P_{1,k}} \nonumber\\
  && \quad +
  \frac{\langle\tilde s_{2,k}^\ast\rangle \tilde x_{2,k}
        g_{2,k}'(|\tilde x_{2,k}|^2/P_{2,k})}{P_{2,k}} \biggr\}
\end{eqnarray}
where the quantities $\langle\tilde s_{1,k}^\ast\rangle$ and
$\langle\tilde s_{2,k}^\ast\rangle $ are mean values of the stochastic
background's Fourier amplitudes at each of the two detector sites.  These both
vanish:
\begin{equation}
  \langle \tilde s_{\{1,2\},k}^\ast\rangle
    = \int d\tilde\vv{s}_1 d\tilde\vv{s}_2
    p_{\text{sb}}(\tilde\vv{s}_1,\tilde\vv{s}_2) s_{\{1,2\},k}^\ast = 0
\end{equation}
since the mean values of the Fourier amplitudes are zero.  Hence, as in
Sec.~\ref{ss:2dwnss} one must look for the locally optimal statistic at the
next order in $\epsilon$.  Taking an additional derivative, one can easily
compute $\Lambda_{(2)}$.  As in Sec.~\ref{ss:2dwnss} this consists of two
types of terms.  For the same reasons as before, we discard from this decision
statistic all the \emph{single detector} terms.  This leaves us with the
following generalized cross-correlation statistic:
\begin{eqnarray}
  \Lambda_{\text{GCC}} &=& 16\int d\tilde\vv{s}_1 d\tilde\vv{s}_2
  p_{\text{sb}}(\tilde\vv{s}_1,\tilde\vv{s}_2) \nonumber\\
  && \times \sum_{k,k'=1}^{[(N-1)/2]}
  \frac{\Re(\tilde s_{1,k}^\ast\tilde x_{1,k})
        g_{1,k}'(|\tilde x_{1,k}|^2/P_{1,k})}{P_{1,k}} \nonumber\\
  &&\qquad \times
  \frac{\Re(\tilde s_{2,k'}^\ast\tilde x_{2,k'})
        g_{2,k'}'(|\tilde x_{2,k'}|^2/P_{2,k'})}{P_{2,k'}}.
\end{eqnarray}
Since the expectation value of the product of the stochastic background at the
two sites is given by
\begin{equation}
  \langle\tilde s_{1,k}\tilde s_{2,k'}\rangle
  = \int d\tilde\vv{s}_1 d\tilde\vv{s}_2
  p_{\text{sb}}(\tilde\vv{s}_1,\tilde\vv{s}_2)
  \tilde s_{1,k} \tilde s_{2,k}^\ast
  = \delta_{kk'} \gamma_k \sigma^2_k
\end{equation}
one obtains the generalized cross-correlation statistic
\begin{eqnarray}
  \Lambda_{\text{GCC}} &=& 16\Re\sum_{k=1}^{[(N-1)/2]}
  \frac{\gamma_k\sigma^2_k\tilde x_{1,k}^\ast\tilde x_{2,k}}{P_{1,k}P_{2,k}}
  \nonumber\\
  &&\quad\times
  g_{1,k}'(|\tilde x_{1,k}|^2/P_{1,k})
  g_{2,k}'(|\tilde x_{2,k}|^2/P_{2,k}).
\end{eqnarray}
If the functions $g'$ are replaced by unity, this reduces to the standard
result for the optimal filter for the case where the detector noise is assumed
to be stationary and Gaussian.  For typical non-Gaussian noise models, the
effect of the $g'$ functions is to exclude those frequency bins in which
$|\tilde x_k|^2/P_k$ is large in either detector.

Our second non-Gaussian noise model is similar to the noise burst model used
in Sec.~\ref{ss:sdcnkw}, generalized to the two detector case.  The composite
PDF for this model is
\begin{eqnarray}
  && p(\vv{x}_1,\vv{x}_2|\epsilon) = (2\pi)^{-N} \nonumber\\
  && \times \{ (1-P_1)(1-P_2)(\det\vv{\Sigma})^{-1}
  \exp(-\frac{1}{2}\bs{\xi}^\dag\cdot\vv{\Sigma}^{-1}\cdot\bs{\xi})
  \nonumber\\
  && \quad + P_1(1-P_2)(\det\vv{\Sigma}_1)^{-1}
  \exp(-\frac{1}{2}\bs{\xi}^\dag\cdot\vv{\Sigma}_1^{-1}\cdot\bs{\xi})
  \nonumber\\
  && \quad + P_2(1-P_1)(\det\vv{\Sigma}_2)^{-1}
  \exp(-\frac{1}{2}\bs{\xi}^\dag\cdot\vv{\Sigma}_2^{-1}\cdot\bs{\xi})
  \nonumber\\
  && \quad + P_1P_2(\det\vv{\Sigma}_{12})^{-1}
  \exp(-\frac{1}{2}\bs{\xi}^\dag\cdot\vv{\Sigma}_{12}^{-1}\cdot\bs{\xi})
  \}
\end{eqnarray}
where $P_1$ and $P_2$ are the probabilities of bursts in detectors 1 and 2.
The matrices $\vv{\Sigma}_1$, $\vv{\Sigma}_2$, and $\vv{\Sigma}_{12}$
represent the correlation matrices when a noise burst is present.  As in
Sec.~\ref{ss:sdcnkw}, a burst effectively changes the noise level for the
detector experiencing the burst.  Thus, if there is a burst in detector 1,
simply replace $\vv{R}_1$ with $\bar\vv{R}_1$ in $\vv{\Sigma}$ to obtain
$\vv{\Sigma}_1$.  Then we find
\begin{equation}
  \vv{\Sigma}_1^{-1} \simeq \left[
  \begin{array}{cc}
    \bar\vv{Q}_1 & -\epsilon^2\bar\vv{Q}_1\cdot\vv{S}\cdot\vv{Q}_2 \\
    -\epsilon^2\vv{Q}_2\cdot\vv{S}\cdot\bar\vv{Q}_1 & \vv{Q}_2
  \end{array}
  \right] + O(\epsilon^4)
\end{equation}
and
\begin{equation}
  \ln\det\vv{\Sigma}_1 \simeq \ln\det\bar\vv{R}_1 + \ln\det\vv{R}_2
    + \tr(\bar\vv{Q}_1\cdot\vv{R}_1) + O(\epsilon^4)
\end{equation}
to first order in $\bar\vv{Q}_1$ and similarly for $\vv{\Sigma}_2$.  We also
have
\begin{equation}
  \vv{\Sigma}_{12}^{-1} \simeq \left[
  \begin{array}{cc} \bar\vv{Q}_1 & \vv{0} \\ \vv{0} & \bar\vv{Q}_2 \end{array}
  \right]
\end{equation}
and
\begin{eqnarray}
  \ln\det\vv{\Sigma}_1 &\simeq& \ln\det\bar\vv{R}_1 + \ln\det\bar\vv{R}_2
  \nonumber\\
  &&\quad  + \tr(\bar\vv{Q}_1\cdot\vv{R}_1) + \tr(\bar\vv{Q}_2\cdot\vv{R}_2)
\end{eqnarray}
to first order in $\bar\vv{Q}_1$ and $\bar\vv{Q}_2$.

We can now compute the locally optimal statistic:
\begin{equation}
  \Lambda_{(2)} \simeq 
    \frac{2\Re(\vv{x}_2^\dag\cdot\vv{Q}_2\cdot\vv{S}\cdot\vv{Q}_1\cdot\vv{x})}%
         {1 + \alpha_1 + \alpha_2 + \alpha_1\alpha_2}
\end{equation}
where
\begin{equation}
  \alpha_1 \simeq \frac{P_1}{1-P_1} \frac{\det\vv{R}_1}{\det\bar\vv{R}_1}
  \exp(\frac{1}{2}\vv{x}_1^\dag\cdot\vv{Q}_1\cdot\vv{x})
\end{equation}
and $\alpha_2$ is given by a similar expression.  Here we have neglected all
$\bar\vv{Q}$ terms.  The terms $\alpha_1$ and $\alpha_2$ detect bursts, and
their role is to suppress $\Lambda_{(2)}$ when a burst is present in either
detector.

\subsection{Estimators}
\label{ss:estimators}



In analyzing experimental data, there are different possible goals.
One goal might be to set an upper limit (with a certain statistical
confidence) on the stochastic background energy density in a particular
frequency band.  Another goal might be to estimate this energy density
in a particular frequency band.

For this latter purpose, there are different possible estimators that might be used.  One standard estimator is the Maximum Likelihood Estimators (MLE).  In this section, we show how this estimator is related to the cross-correlation statistic.

Recall that the probability distribution for the joint detector output is
\begin{eqnarray}
  \ln p(\vv{x}_1,\vv{x}_2|\epsilon) &=&
  \mbox{(terms that don't depend on $\epsilon$)} \nonumber \\
  && +
  \epsilon^2 \vv{x}_2^\dag\cdot\vv{Q}_2\cdot\vv{S}\cdot\vv{Q}_2\cdot\vv{x}_1
  \nonumber\\
  && +
  \frac{1}{2}\epsilon^4\{\tr(\vv{Q}_2\cdot\vv{S}\cdot\vv{Q}_1\cdot\vv{S})
  \nonumber\\
  && \quad -
  \vv{x}_1^\dag\cdot\vv{Q}_1\cdot\vv{S}\cdot\vv{Q}_2\cdot
    \vv{S}\cdot{Q}_1\cdot\vv{x}_1 \nonumber\\
  && \quad -
  \vv{x}_2^\dag\cdot\vv{Q}_2\cdot\vv{S}\cdot\vv{Q}_1\cdot
    \vv{S}\cdot{Q}_2\cdot\vv{x}_2 \}\nonumber\\
  && + O(\epsilon^6).
\end{eqnarray}
Suppose we wish to estimate the strength $\epsilon^2$ of the stochastic
background.  The Maximum Likelihood Estimator is the value
$\epsilon^2_{\text{MLE}}$ for which this probability is maximized:
$[d\ln p(\vv{x}_1,\vv{x}_2|\epsilon)/d\epsilon^2]_{\epsilon^2_{\text{MLE}}}
=0$.  The result is
\begin{equation}
  \epsilon^2_{\text{MLE}} = \eta
  \vv{x}_2^\dag\cdot\vv{Q}_2\cdot\vv{S}\cdot\vv{Q}_2\cdot\vv{x}_1
\end{equation}
where
\begin{eqnarray}
  \eta^{-1} &=& -\tr(\vv{Q}_2\cdot\vv{S}\cdot\vv{Q}_1\cdot\vv{S}) \nonumber\\
  && + \vv{x}_1^\dag\cdot\vv{Q}_1\cdot\vv{S}\cdot\vv{Q}_2\cdot
       \vv{S}\cdot{Q}_1\cdot\vv{x}_1 \nonumber\\
  && + \vv{x}_2^\dag\cdot\vv{Q}_2\cdot\vv{S}\cdot\vv{Q}_1\cdot
       \vv{S}\cdot{Q}_2\cdot\vv{x}_2
\end{eqnarray}
is a measure of how how sensitive the detectors were to the stochastic
background.  Normally $\eta$ will be on the order of unity so
$\epsilon^2_{\text{MLE}}$ is approximately just the cross-correlation
statistic.  However, if the detector were abnormally noisy, then $\eta$ would
be less that unity and the estimate of the stochastic background strength
would be smaller than the cross-correlation statistic would indicate: this is
a correction that compensates for artificially large values of the
cross-correlation statistic due to noise fluctuations.



Another possible estimator is the Bayesian estimator.  In the long measurement, weak signal case this again yields the same result as the MLE estimator.

\section{IMPLEMENTATION/SIMULATIONS}
\label{s:impsim} 

\subsection{Implementation}
\label{ss:implementation}
A nice feature of these techniques is that in practice, they should be
easy to implement. Work by Scott and Whiting \cite{scott99} has shown
that the PDFs of the Fourier amplitudes in different frequency bins
can be easily obtained.  Since the characteristic time-scale for
stochastic background correlation is $\approx 10 \text{\ ms}$, these
can be computed using data-segments with lengths seconds or tens of
seconds. These PDFs can then be used to determine where to truncate or
clip the correlation, frequency-bin by frequency-bin.  Provided that
the instrument's characteristics are stable over periods of minutes or
hours, it should be simple to accumulate sufficient statistics to
determine the PDFs and therefore the truncation or weighting functions
with reasonable accuracy.

In practice, it may also be desirable to ``discard'' a small part of
the ``attainable-in-principle'' correlation in exchange for obtaining
more robust statistics.  For example, on can arbitrarily zero the 1\%
of frequency bins that are the largest number of standard deviations
away from the mean value (for that bin).  Since the dominant
contribution in any bin always comes from the detector noise, this is
only very weakly correlated with the actual stochastic background
signal, and the net effect is to discard just a bit more than 1\% of
the ``in principle'' attainable signal-to-noise ration.  But in
exchange, the detection statistic becomes far less sensitive to
non-Gaussian detector fluctuations.  The precise effects of such
treatment, and the appropriate truncation thresholds, can be easily
determined with Monte Carlo simulations using simulated signals added
into real detector noise.

In searching for a known waveform (e.g., binary inspiral) the methods
are again easily implementable.  Here, since the signal timescale is
less than a minute, the frequency-bin by frequency-bin statistics take
a bit more time to accumulate, and the detector's statistical
properties have got to be stable over a slightly longer time-scale (an
hour, perhaps).  This appears likely.

Since certain non-Gaussian noise features are more likely to appear as
outlier points in the time-domain, and others in the frequency-domain,
a combination of the time- and frequency-domain methods may be
desirable. Unfortunately, if the detector noise is not white, this may
require the removal (vetoing) of entire small sections of time-series
data.  This is easy in the stochastic background case, where only tens
of milliseconds around a glitch need excision.  It may be more
problematic for signals like binary inspiral chirps that have longer
duration.

\subsection{Comparing different statistics}
\label{s:comparing}

In Secs.~\ref{s:deterministic} and \ref{s:stochastic}, 
we derived locally optimal statistics to
search for deterministic and stochastic gravitational wave signals in
the presence of non-Gaussian noise.  These statistics reduce to the
standard matched-filtering and cross-correlation statistics when the
detector noise is Gaussian.  But they are more \emph{robust} (i.e.,
less sensitive to outliers) when the detector noise has non-Gaussian
components.  For both cases, the standard and robust statistics take
(as input) the output of one or more detectors, and return (as output)
a single real number.  But the statistics also depend on the
gravitational wave signal and detector noise models, which are not
directly observable.  Different choices for the signal and noise
models correspond to different statistics, and these different
statistics will in general perform differently given the same detector
output.  In order to compare and evaluate the statistics, we need a
way to quantify their performance.

As mentioned in Sec.~\ref{s:deterministic}, the quality of a test
(i.e., a decision rule based on a particular statistic) is
characterized by its \emph{false alarm} and \emph{false dismissal}
probabilities for a given source.  These are, respectively, the
probability that the test leads us to conclude that a signal is
present, when in fact it is absent ($\epsilon=0$), and the probability
that the test leads us to conclude that a signal is absent, when in
fact it is present ($\epsilon>0$).  These two probabilities (denoted
$\alpha$ and $\beta_\epsilon$) completely specify the long-term
performance of a statistic.  But to rank different tests, we need to
reduce these multi-dimensional error measures to a single figure of
merit.  How we do this depends on the problem we are trying to solve
(see, e.g. \cite{finn98}), but in the context of gravitational wave
detection, it is common to look for a test that minimizes the false
dismissal probability, keeping the false alarm probability less than
or equal to some maximum tolerable value.  This criterion is known as
the \emph{Neyman-Pearson} criterion, and it was used in
Sec.~\ref{s:deterministic} to define the locally optimal statistics.

Thus, to compare the performance of different statistics, we should
plot false dismissal versus false alarm curves for different values of
the signal amplitude $\epsilon$.  The best test (or best statistic) is
the one that has the smallest false dismissal probability
$\beta_{\epsilon}(\alpha)$, for fixed false alarm probability $\alpha$
and fixed signal amplitude $\epsilon$.  Note that since the false
dismissal probability depends on both $\alpha$ and $\epsilon$, it is
possible that the best test for one choice of $(\alpha,\epsilon)$ is
not the best test for a different choice of $(\alpha,\epsilon)$.  Note
also that this method of comparing statistics is different than simply
comparing expected signal-to-noise ratios.  What is important when
determining error rates (and hence the performance of a particular
test) is not the expected value of the statistic, but rather its
probability \emph{distribution}.

\begin{figure}[htb!]
\begin{center}
\epsfig{file=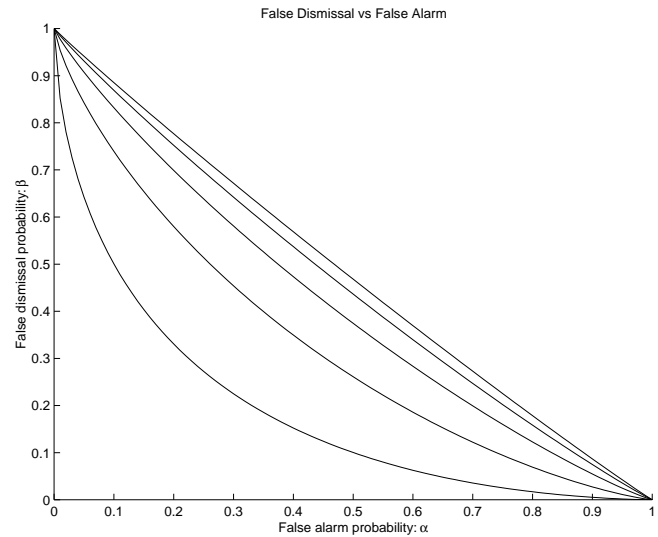,
angle=90,width=3.4in}
\caption{\label{f:typical1}
False dismissal versus false alarm curves for a typical statistic.
Lower curves correspond to larger values of the signal amplitude
$\epsilon$.
}
\end{center}
\end{figure}

For sufficiently simple statistics with sufficiently simple signal and
noise models, it may be possible to analytically calculate the
corresponding false dismissal versus false alarm curves.  But for most
cases of interest, we must resort to \emph{Monte Carlo simulations} to
generate the curves.  This consists of adding simulated signals to
simulated (or real) detector noise, and then processing the resulting
data with a statistic.  For each stretch of data, the statistic
outputs a single number which is then compared to a threshold to
determine if we should claim detection.  Since we know if a signal is
present in the data, we can easily determine the fraction of times
that the decision rule was in error.  In the absence of a signal, this
procedure yields the false alarm probability $\alpha$ as a function of
the threshold $\Lambda_0$.  In the presence of a signal having fixed
amplitude $\epsilon$, we obtain the false dismissal probability
$\beta_\epsilon$, again as a function of the threshold.  If we invert
$\alpha(\Lambda_0)$ for $\Lambda_0=\Lambda_0(\alpha)$, and substitute
this expression back into $\beta_\epsilon(\Lambda_0)$, we obtain the
false dismissal versus false alarm curve $\beta_\epsilon(\alpha)$.  We
can then repeat these steps for a different signal amplitude
$\epsilon'$ to produce a new curve $\beta_{\epsilon'}(\alpha)$.  The
final result will be a set of curves similar to those shown in
Fig.~\ref{f:typical1}.

\begin{figure}[htb!]
\begin{center}
\epsfig{file=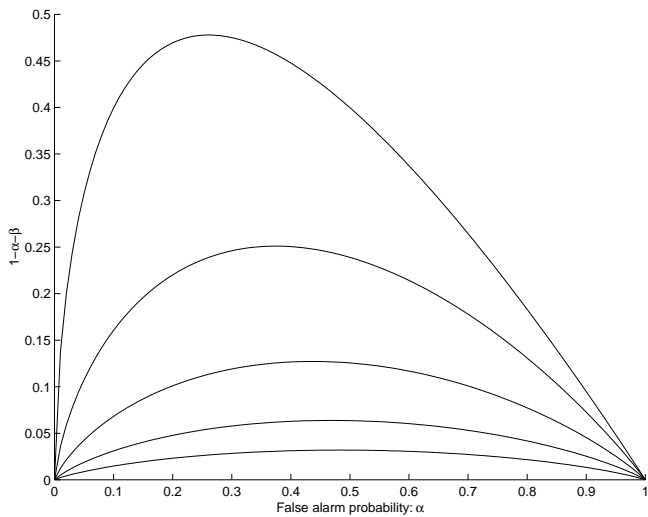,
angle=90,width=3.4in}
\caption{\label{f:typical2}
$1-\alpha-\beta_\epsilon$ versus the false alarm probability $\alpha$
for a typical statistic.
Lower curves correspond to smaller values of the signal amplitude
$\epsilon$.
}
\end{center}
\end{figure}

\begin{figure}[htb!]
\begin{center}
\epsfig{file=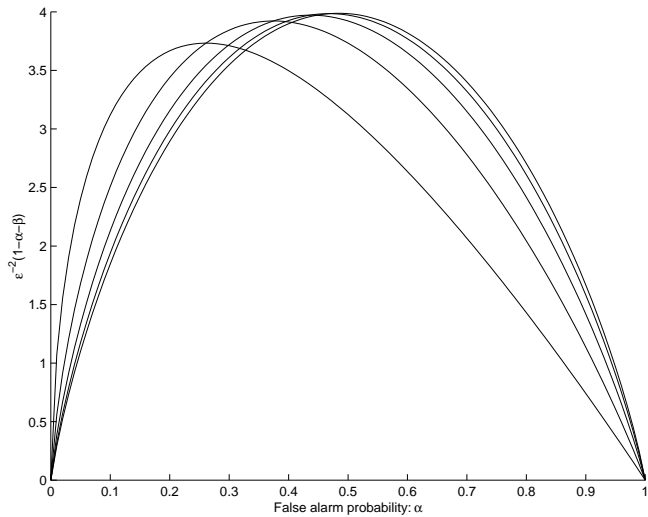,
angle=90,width=3.4in}
\caption{\label{f:typical3}
$\epsilon^{-2}(1-\alpha-\beta_\epsilon)$ versus the false alarm 
probability $\alpha$ for a typical statistic.
Higher curves correspond to smaller values of the signal amplitude
$\epsilon$.
}
\end{center}
\end{figure}

Alternatively, we can plot $1-\alpha-\beta_{\epsilon}$ or
$\epsilon^{-2}(1-\alpha-\beta_{\epsilon})$ versus $\alpha$, as shown
in Figs.~\ref{f:typical2} and \ref{f:typical3}.  Note that the
quantity $1-\alpha-\beta_\epsilon$ is the difference of two
probabilities: $1-\beta_\epsilon$ is the probability that the
statistic exceeds some threshold in the presence of a signal
($\epsilon>0)$, while $\alpha$ is the probability that the statistic
exceeds the same threshold in the absence of a signal (i.e.,
$\epsilon=0$).  Although, Figs.~\ref{f:typical2} and \ref{f:typical3}
contain the same information as the false dismissal versus false alarm
curves (Fig.~\ref{f:typical1}), plotting
$\epsilon^{-2}(1-\alpha-\beta_{\epsilon})$ versus $\alpha$ has the
nice property that, for stochastic signals, the curves have a
well-defined $\epsilon\rightarrow 0$ limit.
%


\subsection{Example}
\label{s:example}

To illustrate how we can compare different statistics using Monte
Carlo simulations, consider the simple case of a search for a white,
Gaussian stochastic background signal using two independent,
identical, coincident and coaligned detectors.  Statistic 1 will be
the standard cross-correlation statistic defined by a white, Gaussian
stochastic background signal and white, Gaussian detector noise.
Statistic 2 will be a locally optimal statistic, also defined by a
white, Gaussian stochastic background signal, but with white,
2-component, mixture Gaussian noise with an arbitrary knee.  We will
assume that we know (a~priori) that the two detectors are identical
and have uncorrelated white noise.  We will not assume, however, that
we know (a~priori) the parameters describing the statistical
properties of the detector noise or the overall amplitude of the
stochastic background signal.  Each statistic will have to internally
estimate the parameters from the detector output, without any other
prior knowledge.

We perform Monte Carlo simulations of the two statistics for the 
following three cases:
\begin{itemize}
\item[(i)] 
uncorrelated, white, Gaussian detector noise with zero mean and unit 
variance.
\item[(ii)] 
uncorrelated, white, 2-component, mixture Gaussian detector noise with 
zero mean, unit variance, $\bar\sigma/\sigma=4$, and $P=1\%$ 
[see Eq.~(\protect\ref{sgs})].
\item[(iii)] 
uncorrelated, white, exponential detector noise with zero mean and 
unit variance [see Eq.~(\protect\ref{e:exp})].
\end{itemize}
The first two simulations test the optimal behavior of the statistics.
Statistic 1 is designed for the data of case (i), and Statistic 2 is
designed for the data of case (ii).  The third simulation tests the
two statistics in a sub-optimal situation, representative of a real
search where we do not know in advance the exact statistical character
of the detector noise.

Details of the Monte Carlo simulations are summarized below:

(i) A single stretch of data consists of $N=1024$ discrete-time
samples.  This $N$ is sufficiently large that the large observation
time approximation is valid.  Since we are considering white noise 
(which has zero correlation length), any $N\gtrsim 100$ would do.

(ii) The simulated stochastic gravitational-wave signal strengths are
$\epsilon^2=.0025$, .005, .010, .020, and .040, where $\epsilon$ is
the ratio of the rms amplitude of the stochastic background signal to
the rms amplitude of the detector noise.  These signal strengths
correspond to signal-to-noise ratios ($\sim\epsilon^2\sqrt{N}$)
ranging from approximately 0.1 to 1 for a single stretch of data.

NOTE: Since a real stochastic background is expected to have a smaller
value of $\epsilon^2$ ($\sim 10^{-4}$), we would need a much longer
observation time to build-up similar signal-to-noise ratios in a real
search.  The purpose of this example, however, is to illustrate how
one can compare two different statistics; it is not meant to simulate
a real ($\gtrsim 4$ month) stochastic background search.

(iii) For all three types of simulated detector noise, the standard
cross-correlation statistic estimates the variance of the noise by
calculating the sample variance of a stretch of detector output equal
to $100N$.  Since the detector output consists in general of signal
plus noise, the estimate of the noise variance gets worse as the
signal amplitude increases.  The sample variance is needed to specify
the white, Gaussian noise model that enters the definition of the
standard cross-correlation statistic [c.f.\ Eq.~\ref{e:gcc}]:
\begin{equation}
{}^{(1)}\Lambda_{\rm GCC}={1\over N}\sum_{j=0}^{N-1}
x_{1,j}x_{2,j}/\sigma^2_1\sigma^2_2\ ,
\label{e:stat1}
\end{equation}
where $\sigma_1^2$ and $\sigma_2^2$ are the estimated variances of 
the noise in detectors 1 and 2, respectively.

(iv) In addition to estimating the variance of the detector noise, the
locally optimal statistic also estimates the variances $\sigma$,
$\bar\sigma$, and breakpoint $x_b$ of the 2-component, mixture
Gaussian model that define this statistic.  It does this by fitting
two straight lines to a $\ln(p(x))$ vs.\ $x^2$ plot obtained from a
histogram of a stretch of detector output, again equal to $100 N$.
Best-fit lines at small $x$ and large $x$, respectively, yield
estimates of $\sigma$ and $\bar\sigma$, while the intersection of the
lines yields an estimate of $x_b$.  Actually, only the breakpoints for
the detector noise are needed to define the following locally optimal
statistic:
\begin{eqnarray}
{}^{(2)}\Lambda_{\rm GCC}={1\over N}\sum_{j=0}^{N-1}&&
x_{1,j}\Theta\left(x_{1b}-|x_{1,j}|\right)\times
\nonumber\\
&&
x_{2,j}\Theta\left(x_{2b}-|x_{2,j}|\right)/\sigma^2_1\sigma^2_2\ ,
\label{e:stat2}
\end{eqnarray}
which is a truncated version of ${}^{(1)}\Lambda_{\rm GCC}$.
(See the discussion of truncation in the previous subsection.)
Here $\Theta(x)$ is the usual step function, which equals 0 if
$x<0$, and equals 1 if $x\ge 0$.

NOTE: In order to handle pure Gaussian noise (which is a pathological
case when one tries to model it as a 2-component, mixture Gaussian
distribution), the locally optimal statistic sets the breakpoint $x_b$
to $\infty$ whenever the estimated slopes at small and large values of
$x$ have a percent difference less than $10\%$.  By doing this, the
locally optimal statistic ${}^{(2)}\Lambda_{\rm GCC}$ effectively 
reduces to the standard cross-correlation statistic 
${}^{(1)}\Lambda_{GCC}$ when the noise is pure Gaussian.

(v) We use $10^5$ trials to generate each false dismissal versus false 
alarm curve.

(vi) The simulations were written in Matlab \cite{MATLAB}.

\begin{figure}[htb!]
\begin{center}
\epsfig{file=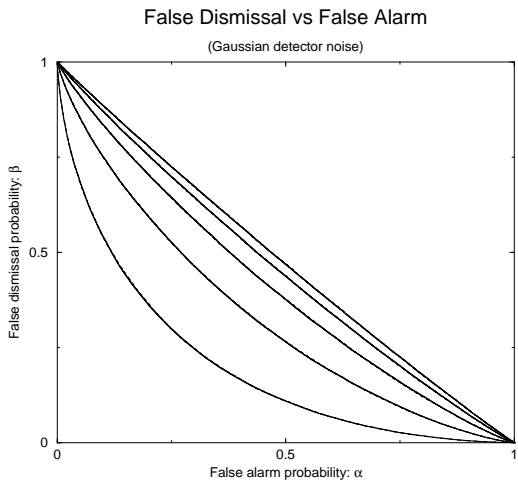,
angle=-90,width=3.4in,bbllx=25pt,bblly=50pt,bburx=590pt,bbury=740pt}
\caption{\label{f:noise1_beta}
False dismissal versus false alarm curves for the standard
cross-correlation and locally optimal statistics for simulated white,
Gaussian detector noise.  The solid lines correspond to the standard
cross-correlation statistic; the dashed lines correspond to the
locally optimal statistic.  The top curve for each statistic has
$\epsilon^2 =.0025$; $\epsilon^2$ increases by a factor of 2 as one
moves to successively lower curves in the graph.  As explained in the
text, the two statistics perform almost identically for this case.
}
\end{center}
\end{figure}

\begin{figure}[htb!]
\begin{center}
\epsfig{file=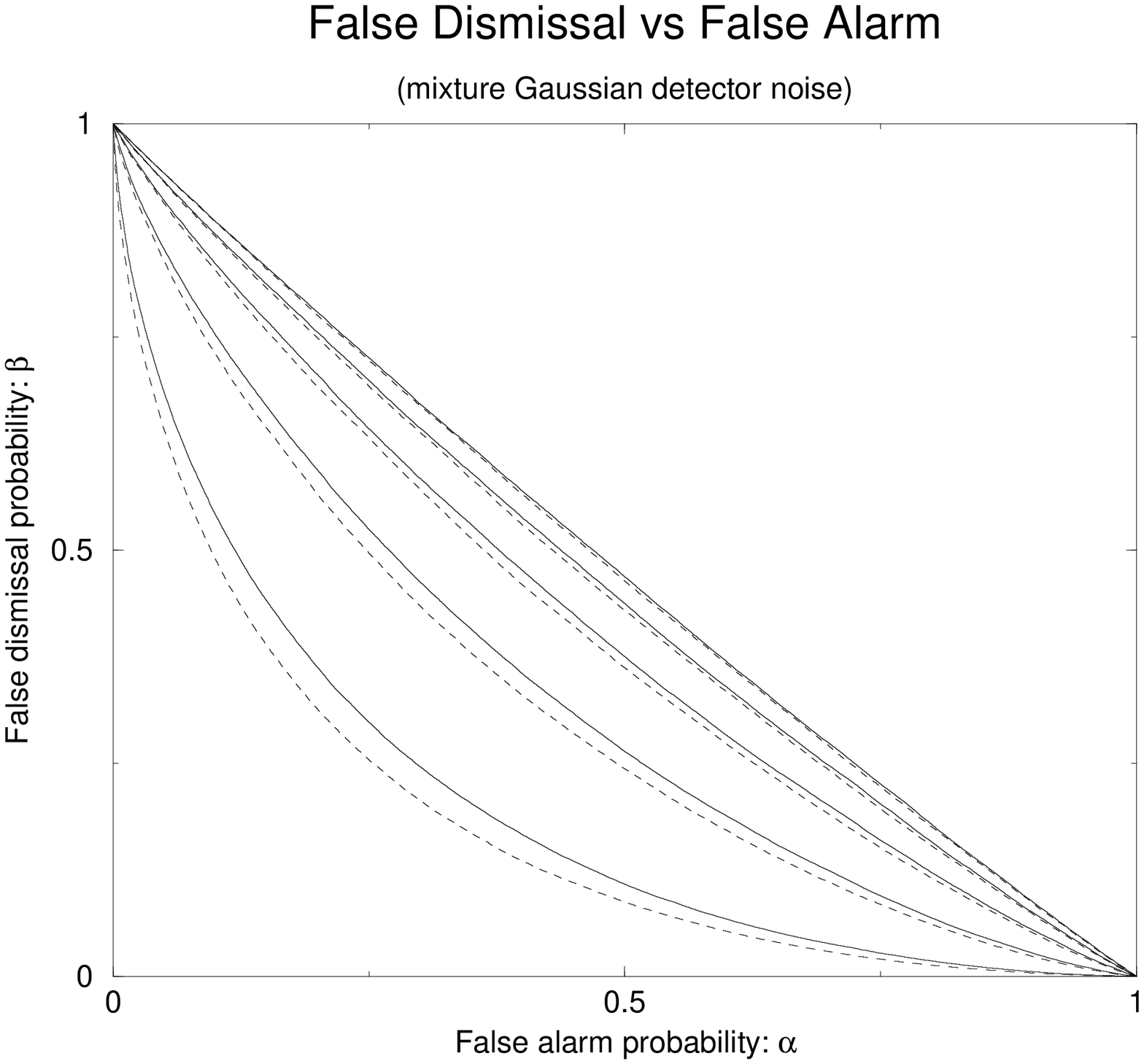,
angle=-90,width=3.4in,bbllx=25pt,bblly=50pt,bburx=590pt,bbury=740pt}
\caption{\label{f:noise2_beta}
False dismissal versus false alarm curves for the standard
cross-correlation and locally optimal statistics for simulated white,
2-component, mixture Gaussian detector noise.  The solid lines
correspond to the standard cross-correlation statistic; the dashed
lines correspond to the locally optimal statistic.  The top curve for
each statistic has $\epsilon^2 =.0025$; $\epsilon^2$ increases by a
factor of 2 as one moves to successively lower curves in the graph.
Since the locally optimally statistic has a lower false dismissal
probability $\beta_{\epsilon}(\alpha)$ for each false alarm
probability $\alpha$ and each signal amplitude $\epsilon$, it is
clearly the better test for this case, as expected.
}
\end{center}
\end{figure}

\begin{figure}[htb!]
\begin{center}
\epsfig{file=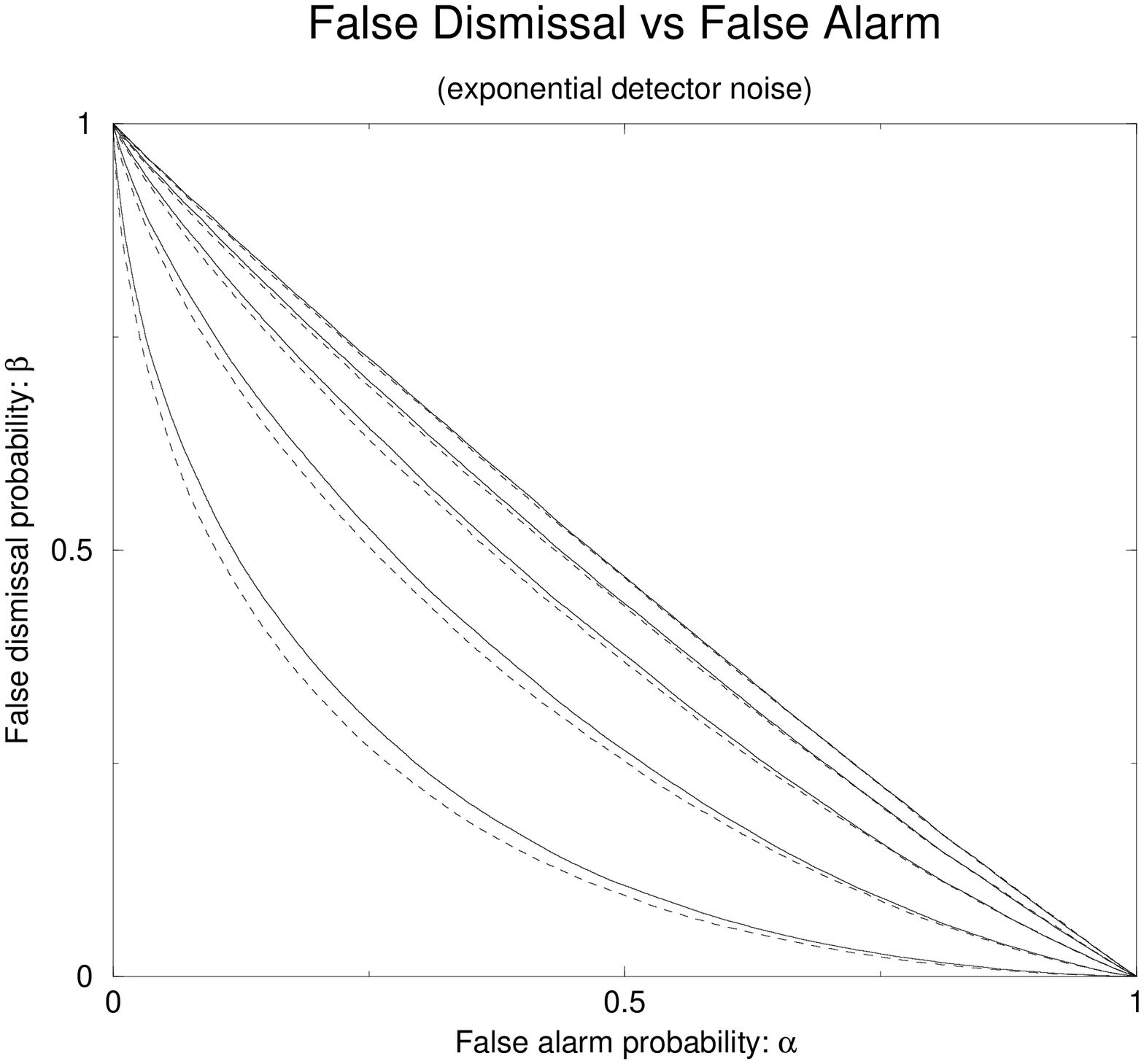,
angle=-90,width=3.4in,bbllx=25pt,bblly=50pt,bburx=590pt,bbury=740pt}
\caption{\label{f:noise3_beta}
False dismissal versus false alarm curves for the standard
cross-correlation and locally optimal statistics for simulated white,
exponential detector noise.  The solid lines correspond to the
standard cross-correlation statistic; the dashed lines correspond to
the locally optimal statistic.  The top curve for each statistic has
$\epsilon^2 =.0025$; $\epsilon^2$ increases by a factor of 2 as one
moves to successively lower curves in the graph.  Since the locally
optimally statistic has a lower false dismissal probability
$\beta_{\epsilon}(\alpha)$ for each false alarm probability $\alpha$
and each signal amplitude $\epsilon$, it is the better test for this
case.
}
\end{center}
\end{figure}

\begin{figure}[htb!]
\begin{center}
\epsfig{file=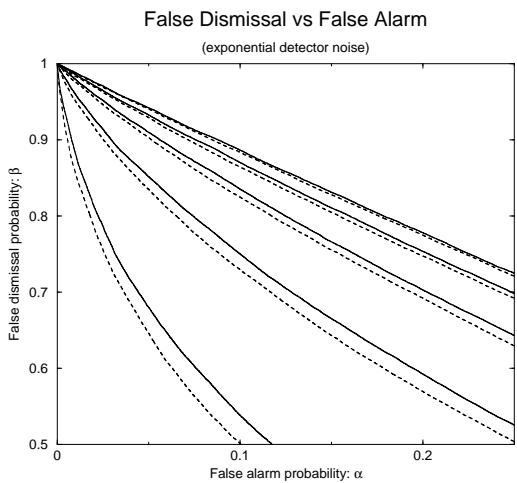,
angle=-90,width=3.4in,bbllx=25pt,bblly=50pt,bburx=590pt,bbury=740pt}
\caption{\label{f:noise3_blowup}
A blow-up of the false dismissal versus false alarm curves from
Fig.~\ref{f:noise3_beta} for small values of the false alarm 
probability $\alpha$.
}
\end{center}
\end{figure}

\begin{figure}[htb!]
\begin{center}
\epsfig{file=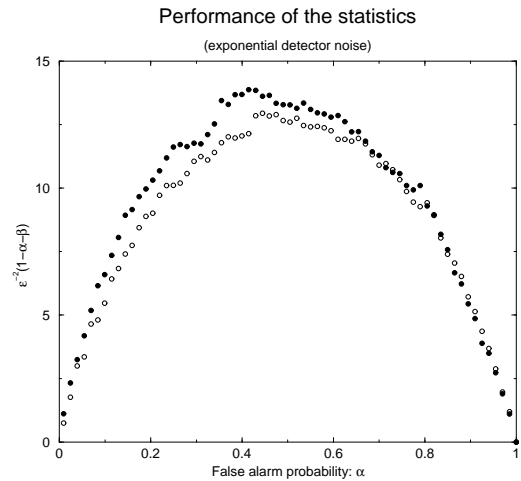,
angle=-90,width=3.4in,bbllx=25pt,bblly=50pt,bburx=590pt,bbury=740pt}
\caption{\label{f:noise3_gamma}
A plot of $\epsilon^{-2}(1-\alpha-\beta_\epsilon)$ versus the false
alarm probability $\alpha$ for the standard cross-correlation and
locally optimal statistics for simulated white, exponential detector
noise in the weak signal limit (small $\epsilon$).  The top curve
(filled circles) corresponds to the locally optimal statistic; the
lower curve (open circles) corresponds to the standard
cross-correlation statistic.  The difference between the performance
of the two statistics in the small signal limit is more apparent in
this plot (cf. Fig.~\ref{f:noise3_beta}).  Since the locally optimally
statistic has a larger value of
$\epsilon^{-2}(1-\alpha-\beta_\epsilon)$ for each false alarm
probability $\alpha$, it is the better test for this case.
}
\end{center}
\end{figure}

The results of the simulation are shown in 
Figs.~\ref{f:noise1_beta}-\ref{f:noise3_gamma}.

As noted in (iv) above, our implementation of the locally optimal 
statistic reduces to the standard cross-correlation statistic when the 
detector noise is pure Gaussian.  That is why the false dismissal 
versus false alarm curves for the two statistics are effectively
identical in Fig.~\ref{f:noise1_beta}.

From Fig.~\ref{f:noise2_beta} we see that the locally optimal statistic
performs better than the standard cross-correlation statistic when the
simulated detector noise is mixture Gaussian.  For each value of the
stochastic signal strength $\epsilon^2$ and for each false alarm
probability $\alpha$, the false dismissal probability
$\beta_\epsilon(\alpha)$ for the locally optimal statistic is less
than that for the standard cross-correlation statistic.  This is as
expected, since the locally optimal statistic was constructed
precisely to handle mixture Gaussian noise.

Finally, from Figs.~\ref{f:noise3_beta}-\ref{f:noise3_gamma} we see
that the locally optimal statistic also performs better than the
standard cross-correlation statistic when the simulated detector noise
has an exponential distribution.  The difference in performance
between the two statistics for this case is less than that for mixture
Gaussian noise, but it is still noticeable.
(Figure~\ref{f:noise3_blowup} focuses attention on the false
dismissal versus false alarm curves for small values of the false
alarm probability, while Fig.~\ref{f:noise3_gamma} is a plot of
$\epsilon^{-2}(1-\alpha-\beta_\epsilon)$ versus $\alpha$, which
highlights the difference between the two statistics in the small
signal limit.)  This behavior is again as expected, since a locally
optimal statistic is constructed to be less sensitive to the tails of a
non-Gaussian distribution.

\section{CONCLUSION}
\label{s:conclude}
In this paper, we have constructed a replacement for the standard linear
matched filter estimators used for gravitational wave detection. The
replacements are more robust because they are less susceptible to corruption
by non-Gaussian detector noise.

We have explicitly illustrated the locally optimal detection strategies for a
variety of different noise PDFs, and for two different detection problems
(single detector known waveform, and two-detector stochastic background).  In
all cases, the optimal strategy is similar to the one for Gaussian noise
except that data samples that lie outside the central part of the distribution
(the outliers) are excluded from the sums which form the estimators.

We believe that for the future generation of sensitive gravitational wave
detectors, these strategies may be easily implemented and offer an improvement
on the existing matched filter algorithms.

\acknowledgments
This research was supported in part by NSF grants PHY-9728704,
PHY-9722189, PHY-9981795, PHY-0071028, NASA grant NASA-JPL 961298, the
Sloan Foundation, and by the Max Planck Society (Albert Einstein
Institute, Potsdam).


\begin{thebibliography}{}

\bibitem{science92} A. Abramovici, W. E. Althouse, R. W. P. Drever,
Y. G{\"u}rsel, S. Kawamura, F. J. Raab, D. Shoemaker, L. Sievers,
R. E. Spero, K. S. Thorne, R. E. Vogt, R. Weiss, S. E. Whitcomb, and
M. E. Zucker, Science \textbf{256}, 325 (1992).

\bibitem{physicstoday} Barry C. Barish and Rainer Weiss, Phys. Today,
October (1999).

\bibitem{virgo} C. Bradaschia et al., Nucl. Instrum. Methods
\textbf{A289}, 518 (1990); also in \textit{Gravitation 1990}, Proceedings
of the Banff Summer Institute, Banff, Alberta, 1990, edited by R. Mann
and P. Wesson (World Scientific, Singapore, 1991).

\bibitem{geo600} K.~Danzmann et al., in \textit{Gravitational Wave
Experiments}, proceedings of the Edoardo Amaldi Conference, edited by 
E. Coccia, G. Pizzella, and F. Ronga (World Scientific, Singapore 1995),
p. 100.

\bibitem{tama300} K.~Tsubono, \textit{ibid}. p. 112.

\bibitem{barsinoperation} I. S. Heng, D. G. Blair, E. N. Ivanov, M. E. Tobar,
Phys. Lett. A \textbf{218}, 190 (1996); E. Mauceli et al., 1996
Phys. Rev. D \textbf{54}, 1264 (1996); P. Astone et al., \textit{ibid}.
\textbf{47}, 362 (1993); M. Cerdonio et al., in \textit{Gravitational Wave
Experiments}, proceedings of the Edoardo Amaldi Conference, edited by 
E. Coccia, G. Pizzella, and F. Ronga (World Scientific, Singapore 1995),
p. 176; E. Coccia, \textit{ibid}., p. 161.

\bibitem{plannedinstruments} Instruments currently in the planning and
proposal stage include the second-generation LIGO detector (LIGO-II), an
advanced European detector, and a 3~km scale-up of the TAMA-300 detector.

\bibitem{300years} K. S. Thorne, in \textit{300 Years of Gravitation},
edited by S. W. Hawking and W. Israel (Cambridge University Press,
Cambridge, England, 1987), pp. 330--458.

\bibitem{exptwith40m} Simple experiments carried out by Allen and Romano using
the GRASP data analysis package 
\texttt{http:\slash\slash www.lsc-group.phys.uwm.edu} indicate that the large
non-Gaussian fluctuations present in the data from the LIGO 40-m prototype
created large biases in the cross-correlation estimators. The 1994 data stream
was split into two parts, and each was assumed to come from an independent
detector.

\bibitem{kassam}
Saleem A. Kassam, \textit{Signal detection in non-Gaussian noise},
(Springer-Verlag, New York, 1988).

\bibitem{creighton99}
Jolien D. E. Creighton, Phys. Rev. D \textbf{60}, 021101 (1999).

\bibitem{allen96}
B. Allen, in \textit{Proceedings of the Les Houches School on Astrophysical
Sources of Gravitational Waves}, edited by Jean-Alain Marck and Jean-Pierre
Lasota (Cambridge University Press, Cambridge, England, 1997).

\bibitem{flanagan93} 
\'E. \'E. Flanagan, Phys. Rev. D \textbf{48}, 2389 (1993).  Note that the
second term on the RHS of Eq. (B6) should read $-10 j_1(\alpha)$ rather than
$-2 j_1(\alpha)$, that the sliding delay function shown in Fig.~2 is
incorrect, and that the right hand side of Eq. (2.8) should be divided by
$\pi$.

\bibitem{allen97}
B. Allen and J. Romano, Phys. Rev. D \textbf{59}, 102001 (1999).

\bibitem{allenflanaganpapa}
Bruce Allen, \'E. \'E. Flanagan, and Maria Alessandra Papa,
Phys. Rev. D \textbf{61}, 024024 (2000).

\bibitem{scott99} Susan Scott and Bernard Whiting, work reported at the LIGO
Scientific Collaboration Meetings in August 1999, and March 2000, and at the
TAMA Meeting in October 1999.

\bibitem{finn98} L. S. Finn, (unpublished).

\bibitem{MATLAB} 
Matlab: \texttt{http:\slash\slash www.mathworks.com}.
The matlab scripts that we used to generate the Monte Carlo simulations 
are available at:
\texttt{http:\slash\slash feynman.utb.edu\slash$\sim$joe\slash research\slash 
stochastic\slash algorithms\slash robust\slash simulations\slash code\slash}.

\end{thebibliography}
\end{document}